\documentclass[%
 reprint,
 superscriptaddress,
 amsmath,amssymb,
 prl,
 nofootinbib,
]{revtex4-2}

\usepackage{graphicx}
\usepackage{bm}
\newcommand{\Tr}{\operatorname{Tr}}
\newcommand{\tr}{\operatorname{tr}}
\newcommand{\Real}{\operatorname{Re}}
\newcommand{\Imag}{\operatorname{Im}}
\usepackage[colorlinks,linkcolor=blue,urlcolor=blue,anchorcolor=blue,citecolor=blue]{hyperref}

\AtBeginDocument{%
  \ifx\bibsep\undefined\else\setlength{\bibsep}{0pt}\fi
}

\newcommand{\Davg}[1]{\Delta\langle #1\rangle}
\newcommand{\LL}{\hat{\mathcal{L}}}   
\newcommand{\WW}{\mathcal{W}}
\newcommand{\DD}{\hat{\mathcal{D}}}

\begin{document}

\preprint{APS/123-QED}

\title{Quantum Data Ruler}

\author{Wangjun Lu}
\email{wjlu1227@zju.edu.cn}
\affiliation{Department of Maths and Physics, Hunan Institute of Engineering, Xiangtan 411104, China}

\date{\today}

\begin{abstract}
A ruler reads an unknown length against a calibrated scale. The quantum data
ruler reads a matrix element of an unknown state against the signal of a known
reference, from their ratio. It is exact for any Hamiltonian and any time, since
transition operators change by a state-dependent constant times one function of
time, removing the vanishing-signal limit. The readout is
unbiased under dephasing and amplitude damping, and recovers the coherent
ergotropy of a quantum battery that two-point measurement erases. The structure survives the
classical limit as action--angle Fourier modes.\end{abstract}

\maketitle

\section{Introduction}

Measuring an observable couples the acquisition of information to the
disturbance of the state \cite{Bohr1928-BOHTQP,von2018mathematical,RevModPhys.75.715,schlosshauer2007decoherence}.
Standard tomography is destructive, requiring many copies and a complete set of
settings \cite{leonhardt1997measuring,RevModPhys.81.299}. The resulting
information--disturbance trade-off has been characterized quantitatively
\cite{PhysRevA.53.2038,PhysRevLett.86.1366,PhysRevLett.100.210504}, and much
work aims to mitigate the disturbance: quantum non-demolition measurement
\cite{braginsky1980quantum,grangier1998quantum}, weak measurements and weak
values \cite{PhysRevLett.60.1351,PhysRevLett.66.1107,jordan2020quantum},
direct measurement of wavefunctions
\cite{lundeen2011direct,PhysRevLett.108.070402}, and quantum state holography
\cite{PhysRevLett.80.1418,weinacht1999controlling}.

In metrology one never measures a length directly but compares it against a
calibrated scale; the same holds for quantum states. Hats denote operators
throughout. Let an unknown initial state
$\hat\rho_0$ and a known reference state $\hat\rho_r$ evolve under the same
Hamiltonian $\hat H$ for a time $t$, let $\hat\rho(t)=e^{-i\hat Ht}\hat\rho_0e^{i\hat Ht}$
be the evolved unknown state, and let the change
$\Davg{\hat A}(t)=\Tr[\hat A\hat\rho(t)]-\Tr[\hat A\hat\rho_0]$ of the
expectation value of a time-independent observable $\hat A$ be measured in each
run. Because $\hat\rho_r$ is known, the meaning of its signal is known: the
reference is the data ruler, and the ratio of the two signals encodes a matrix
element of $\hat\rho_0$. Unlike in holographic schemes \cite{PhysRevLett.80.1418}, the reference need not
join the target dynamics.

Four facts make the construction significant. First, the comparison is
exact at arbitrary time and for an arbitrary Hamiltonian
(Theorem~1). An indirect scheme infers an initial observable from the instantaneous
change of another \cite{Lu2022InstMeas,PhysRevA.104.043706}, facing a vanishing
signal as $t\to0$ --- L'H\^opital's rule on a quantity vanishing there; a known
reference cancels the entire time dependence, not merely its leading term,
turning that signal into one of order unity. Second, every pair of levels
$(m,n)$ carries an embedded spin-1/2 algebra (Observation~1): the ruler reads two
in-plane components of its Bloch vector, so the trade-off becomes geometric and
the precision acquires a bound, Eq.~\eqref{eq:uncertainty}, set by the population
imbalance $p_m-p_n$ --- $p_n=\langle E_n|\hat\rho_0|E_n\rangle$ being the initial
population of level $n$ --- which is also the single switch for the incompatibility
of the joint readout (Theorem~2). Third, the ratio is unbiased under any noise
satisfying a spectral criterion (Theorem~3) --- phase and amplitude damping
included, of unknown strength --- and requires no knowledge of the levels of
$\hat H$. Fourth, in the classical limit the eigen-operators become
action--angle Fourier modes and comparing two ensembles is again exact in time;
the correspondence is one-way (Sec.~S7 \cite{supp}).

\section{The ruler identity}

Let $\hat H$ be an arbitrary Hermitian Hamiltonian on a $d$-dimensional space,
$\hat\rho_0$ an arbitrary initial state ($\hbar=1$), and
$\hat\rho(t)=e^{-i\hat Ht}\hat\rho_0e^{i\hat Ht}$. The change of the expectation
value of a time-independent observable $\hat A$ obeys the exact identity
\begin{equation}
\Davg{\hat A}(t,\hat\rho_0)=\operatorname{Tr}\!\big[(e^{i\hat Ht}\hat Ae^{-i\hat Ht}-\hat A)\hat\rho_0\big]
=\operatorname{Tr}\!\big[\LL_t(\hat A)\hat\rho_0\big],
\label{eq:identity}
\end{equation}
with $\LL_t(\hat A):=e^{i\hat Ht}\hat Ae^{-i\hat Ht}-\hat A$. No small-time expansion is involved: the
readout is a linear functional of $\hat\rho_0$ whose kernel $\LL_t(\hat A)$ is known
once $\hat H$ is known, or calibrated by a reference.
Expanding the exponential recovers the instantaneous scheme of
Refs.~\cite{Lu2022InstMeas,PhysRevA.104.043706} as its leading order. The subscript $t$ marks this finite-time map; for Hamiltonian dynamics
$\LL_t=e^{\LL^\dagger t}-1$, $\LL^\dagger$ the adjoint under $\Tr(\hat X^\dagger\hat Y)$
(Sec.~\ref{sec:noise}).

\begin{figure}[t]
\includegraphics[width=1.0\columnwidth]{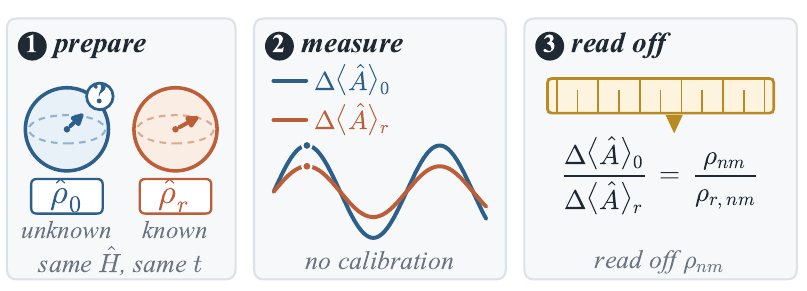}
\caption{Quantum data ruler. An unknown state $\hat\rho_0$ and a known reference
$\hat\rho_r$ evolve under the same $\hat H$; the ratio of the measured changes
$\Delta\langle\hat A\rangle(t)$ equals the ratio of the initial matrix elements,
so the unknown is read off against the known scale.}
\label{fig:schematic}
\end{figure}

\noindent\textbf{Theorem 1 (ruler identity).}
Let $\{|E_n\rangle\}_{n=1}^{d}$ be an eigenbasis of $\hat H$,
$\hat H|E_n\rangle=E_n|E_n\rangle$, $\omega_{mn}=E_m-E_n\neq0$, and
$\hat A_{mn}=|E_m\rangle\langle E_n|$. Then
$[\hat H,\hat A_{mn}]=\omega_{mn}\hat A_{mn}$ and
\begin{equation}
\Davg{\hat A_{mn}}(t)=\big(e^{i\omega_{mn}t}-1\big)\rho_{nm},
\label{eq:transition}
\end{equation}
where $\rho_{nm}:=\langle E_n|\hat\rho_0|E_m\rangle$. Hence for any unknown
$\hat\rho_0$ and any known reference $\hat\rho_r$ with
$\rho_{r,nm}:=\langle E_n|\hat\rho_r|E_m\rangle\neq0$,
\begin{equation}
\frac{\Davg{\hat A_{mn}}(t,\hat\rho_0)}{\Davg{\hat A_{mn}}(t,\hat\rho_r)}
=\frac{\rho_{nm}}{\rho_{r,nm}},\qquad \forall t\ (e^{i\omega_{mn}t}\neq1).
\label{eq:ratio}
\end{equation}

\noindent\textbf{Proof.} $[\hat H,|E_m\rangle\langle E_n|]=\omega_{mn}|E_m\rangle\langle E_n|$,
so the Heisenberg equation $d\hat A_{mn}/dt=i[\hat H,\hat A_{mn}]=i\omega_{mn}\hat A_{mn}$ gives
$\hat A_{mn}(t)=e^{i\omega_{mn}t}\hat A_{mn}$, whence Eq.~\eqref{eq:transition}; dividing
the unknown and reference versions cancels the common factor
$e^{i\omega_{mn}t}-1$.

Equation~\eqref{eq:ratio} is the quantum data ruler (Fig.~\ref{fig:schematic}):
the reference removes the
time dependence, the transition frequency $\omega_{mn}$, and any coupling
constant at once, giving $\rho_{nm}=\rho_{r,nm}R(t)$ with
$R(t)=\Davg{\hat A_{mn}}(t,\hat\rho_0)/\Davg{\hat A_{mn}}(t,\hat\rho_r)$ at any $t$. The signal
$|\Davg{\hat A_{mn}}(t)|=2|\rho_{nm}||\sin(\omega_{mn}t/2)|$ reaches $2|\rho_{nm}|$
at $t=\pi/|\omega_{mn}|$; there is no vanishing-signal limit. In practice one observable $\hat A=\sum_{mn}a_{mn}\hat A_{mn}$
suffices, since $a_{mn}$ and $e^{i\omega_{mn}t}-1$ are common to both states: the
ratio of the fitted amplitudes at each line returns Eq.~\eqref{eq:ratio} with
neither eigenbasis nor levels known \cite{supp}. Theorem~1 holds for any Hamiltonian, integrable or
chaotic.

\section{Level-pair qubit}

The Hermitian combinations $\hat X_{mn}=\hat A_{mn}+\hat A_{nm}$,
$\hat Y_{mn}=i(\hat A_{nm}-\hat A_{mn})$, and
$\hat Z_{mn}=|E_m\rangle\langle E_m|-|E_n\rangle\langle E_n|$ are observables.

\noindent\textbf{Observation 1 (level-pair qubit).}
For any $\hat H$ and any pair $(m,n)$, $\{\hat X_{mn},\hat Y_{mn},\hat Z_{mn}\}$ satisfy the Pauli
algebra
\begin{equation}
[\hat X_{mn},\hat Y_{mn}]=2i\hat Z_{mn},\quad\text{and cyclically}.
\label{eq:su2}
\end{equation}

\noindent\textbf{Proof.} With $\hat A=|E_m\rangle\langle E_n|$ one has
$[\hat A,\hat A^\dagger]=\hat Z_{mn}$ and $[\hat A,\hat Z_{mn}]=-2\hat A$; the three relations follow by direct
evaluation. 

This is not new: it is the standard $\mathfrak{su}(2)$ triple of a pair of
roots \cite{FultonHarris1991}, the Gell-Mann basis in $d\times d$ form and the
pseudo-spin of Ref.~\cite{Feynman1957}; qudits have likewise been represented by
one Bloch vector per level pair \cite{Kurzynski2011,LopezSaldivar2019}. What matters
is that the ruler reads two components of that spin,
$\rho_{nm}=\frac12\langle \hat X_{mn}\rangle+\frac{i}{2}\langle \hat Y_{mn}\rangle$, with
$\langle \hat Z_{mn}\rangle=p_m-p_n$, where
$\langle\hat O\rangle:=\Tr(\hat O\hat\rho_0)$: the single-frequency readout inverts
the $2\times2$ map from $(\Real\rho_{nm},\Imag\rho_{nm})$ to
$(\Davg{\hat X_{mn}},\Davg{\hat Y_{mn}})$ whenever $\cos\omega_{mn}t\neq1$, giving both
components at finite time \cite{supp}.

Non-commutativity bounds the precision. The Robertson uncertainty
relation \cite{Robertson1929} --- the general form of Heisenberg's,
$\Delta\hat B\,\Delta\hat C\geq\frac12|\langle[\hat B,\hat C]\rangle|$
with $\Delta\hat B:=\sqrt{\langle\hat B^2\rangle-\langle\hat B\rangle^2}$ --- and
Eq.~\eqref{eq:su2} give
\begin{equation}
 \Delta \hat X_{mn}\,\Delta \hat Y_{mn}\geq |p_m-p_n|
\label{eq:uncertainty}
\end{equation}
the floor being set by the population imbalance and closing at $p_m=p_n$. That
is only a bound. Normalising the pair state, $P=p_m+p_n$, $\mathbf r=(x,y,r_z)$
with $x+iy=2\rho_{nm}/P$, $r_z=(p_m-p_n)/P$ --- the Bloch vector of the
normalised $2\times2$ block $\hat\varrho$ of $\hat\rho_0$ --- the SLD Fisher
matrix $F$, with entries $F_{ij}=\Real\Tr(\hat\varrho\hat L_i\hat L_j)$ for
$i,j\in\{x,y\}$, gives $\tr F^{-1}=2-(x^2+y^2)/(1-r_z^2)$ and
$|\Tr\hat\varrho[\hat L_x,\hat L_y]|=2|r_z|$, where the SLD operators $\hat L_x,\hat L_y$
are defined by $\partial_x\hat\varrho=\frac12(\hat\varrho\hat L_x+\hat L_x\hat\varrho)$
and likewise for $y$; attainability is settled by

\noindent\textbf{Theorem 2 (incompatibility switch).}
Let $\mathcal{G}=C^{\rm H}-\tr F^{-1}$ be the Holevo--SLD gap at unit weight for
estimating $x,y$ jointly, where $C^{\rm H}$ is the Holevo bound, and
$s=1-|\mathbf r|^2>0$. Then $\mathcal{G}=0$ if
$p_m=p_n$, while for $p_m\neq p_n$
\begin{equation}
\mathcal{G}\ \geq\ \frac{r_z^2\,s^2}{1-r_z^2}\ >\ 0 .
\label{eq:switch}
\end{equation}

\noindent\textbf{Proof.} The two-parameter qubit Holevo bound reduces to a
minimisation over two reals \cite{Suzuki2016}; at $r_z=0$ its minimum equals
$\tr F^{-1}$, and for $r_z\neq0$ Cauchy--Schwarz reduces it to one variable whose
minimum is the right-hand side of Eq.~\eqref{eq:switch} \cite{supp}. 

Population imbalance is thus the single switch for the Robertson floor and for
the incompatibility blocking a joint readout: at $p_m=p_n$ the bounds coincide
and the SLD Cram\'er--Rao bound is asymptotically attainable. Two layers must
not be conflated: Eq.~\eqref{eq:switch} is asymptotic and may use collective
measurements, whereas a single copy pays a further state-independent
factor $\approx2$ ($1.7$--$2.0$ numerically) \cite{supp}.

\noindent\textbf{Trade-off.} For a pure pair state ($|\mathbf r|=1$) the signal
$S_{mn}=2|\rho_{nm}|=\sqrt{P^2-(p_m-p_n)^2}$ is largest at $p_m=p_n$, exactly where
the floor \eqref{eq:uncertainty} closes \cite{supp}.

\section{Noise-immune readout}
\label{sec:noise}

The ratio form makes the ruler robust against the noise that most damages a
coherence readout.

\noindent\textbf{Theorem 3 (noise immunity).}
Let the dynamics be linear and state-independent, $\hat\rho(t)=e^{\LL t}\hat\rho_0$,
with the Lindblad generator
$\LL(\hat\rho)=-i[\hat H,\hat\rho]+\sum_\alpha\big(\hat L_\alpha\hat\rho\hat L_\alpha^\dagger-\frac12\{\hat L_\alpha^\dagger\hat L_\alpha,\hat\rho\}\big)$.
Then Eq.~\eqref{eq:ratio} holds for all $t$ if and only if each transition
operator is an eigen-operator of the adjoint generator,
$\LL^\dagger(\hat A_{mn})=\lambda_{mn}\hat A_{mn}$. For a Lindbladian with jump operators
$\hat L_\alpha$ this is guaranteed for every pair at once by
\begin{equation}
[\hat L_\alpha^\dagger \hat L_\alpha,\hat H]=0,\quad
\hat L_\alpha\ \text{diagonal or}\ \propto|E_j\rangle\langle E_k|,\ j\neq k,
\label{eq:criterion}
\end{equation}
and then $\lambda_{mn}=i\omega_{mn}+\sum_\alpha\big[c^{(\alpha)*}_{mm}
c^{(\alpha)}_{nn}-\tfrac12(\Gamma_m+\Gamma_n)\big]$, with
$c^{(\alpha)}_{jk}=\langle E_j|\hat L_\alpha|E_k\rangle$ and
$\Gamma_m=\sum_\alpha(\hat L_\alpha^\dagger \hat L_\alpha)_{mm}$.

\noindent\textbf{Proof.} With $\hat K=\hat L^\dagger \hat L$, $\tfrac12\{\hat K,\hat A_{mn}\}\propto
\hat A_{mn}$ for every pair iff $\hat K$ is diagonal in the energy basis, i.e.\
$[\hat K,\hat H]=0$; and $\hat L^\dagger \hat A_{mn}\hat L=\sum_{kk'}c^*_{mk}c_{nk'}|E_k\rangle\langle
E_{k'}|$, $c_{jk}=\langle E_j|\hat L|E_k\rangle$, which is $\propto \hat A_{mn}$ for
diagonal $\hat L$ and vanishes for $\hat L\propto|E_j\rangle\langle E_k|$, $j\neq k$.
Hence $\LL^\dagger(\hat A_{mn})\propto \hat A_{mn}$, so
$\Davg{\hat A_{mn}}(t,\hat\rho_0)=\rho_{nm}(e^{\lambda_{mn}t}-1)$ and the
state-independent bracket cancels in the ratio. 

Equation~\eqref{eq:criterion} covers phase damping ($\hat L_\alpha$ diagonal) and
spontaneous emission ($\hat L_\alpha=|E_j\rangle\langle E_k|$) alike, i.e.\ $T_1$ as
well as $T_2$. It fails for one coherent loss channel serving several
transitions, such as cavity decay $\hat L=\sqrt\kappa\,\hat a$: then
$\hat L^\dagger \hat L=\kappa\hat n$, not diagonal on hybridised eigenstates. The bias is
$O(\kappa/\Delta f_{\min})$, $\Delta f_{\min}$ the smallest spacing between the
sectors the channel couples, an order of magnitude below the shot-noise floor on
the platform of Sec.~S11 \cite{supp}. Local relaxation
$\hat L_i=\sqrt{\gamma_1}\hat\sigma_i^-$ also falls outside it, but costs $\leq0.3\%$ of
the coherent ergotropy at $T_1=20\ \mu$s (Sec.~S11.5). Three caveats: (i) the two states must share the noise; (ii) the theorem
removes a bias, not a loss of resolution; (iii) reading all $d^2$ elements costs
$O(d^2)$ settings, as in tomography.

\section{Quantum--classical correspondence}

The eigen-operator structure is not a peculiarity of quantum mechanics
(Sec.~S7 \cite{supp}); hats mark operators, so the classical quantities of this
section carry none. For an integrable system the action--angle variables
$(I,\theta)$ give $H=H(I)$, and the Fourier modes $e^{ik\cdot\theta}$ obey
$\{e^{ik\cdot\theta},H\}=i(k\cdot\omega)e^{ik\cdot\theta}$, with $\{\,,\,\}$ the Poisson
bracket, $k$ an integer wave vector and $\omega=\partial H/\partial I$; this is the
exact analogue of $[\hat H,\hat A_{mn}]=\omega_{mn}\hat A_{mn}$. Each mode is a classical
eigen-operator, so the ratio of two ensembles' changes again cancels the time
dependence --- a classical data ruler. Quantum mechanics keeps the structure and
drops the hypothesis: spectral decomposition exists for every Hamiltonian but
action--angle variables only for integrable ones, so the classical ruler is
confined to integrable dynamics while Eq.~\eqref{eq:ratio} is universal. The
correspondence is one-way in the other direction too: $\cos k\theta$ and
$\sin k\theta$ have vanishing Poisson bracket, so neither the Robertson floor nor
back-action survives the classical limit.

\section{Application: coherence in the energy eigenbasis}
\label{sec:application}

The two properties above --- no energy levels needed, unbiased under unknown
dephasing --- address a problem blocked by a no-go theorem. In quantum
thermodynamics the work on a driven system is accessed through the two-point
measurement (TPM) scheme, whose first projection replaces $\hat\rho_0$ by its
energy-dephased version
$\DD(\hat\rho_0):=\sum_n p_n|E_n\rangle\langle E_n|$ (the map deleting all
off-diagonal elements in the energy basis), erasing the coherence that drives the process; no
universal positive work distribution reproduces TPM for incoherent states while
giving the correct mean for coherent ones \cite{PerarnauLlobet2017}. The same
erasure limits quantum batteries: the coherent part of the ergotropy, which
decays slowest and sets the optimal switch-off time \cite{Zhang2026QB}, is
measured only by destroying coherence in a control run \cite{Streltsov2017}.

The ruler removes that control run: the coherent
part of any energetic quantity is fixed by the off-diagonal elements
$\rho_{nm}$, which Theorem~1 reads without projecting $\hat\rho_0$ or knowing the
levels of $\hat H$, while Theorem~3 keeps it unbiased. For the ergotropy
$\WW(\hat\rho)=\Tr(\hat\rho \hat H)-\sum_k\lambda_k^\downarrow E_k^\uparrow$, with
eigenvalues $\lambda_k^\downarrow$ of $\hat\rho$ sorted downwards and $E_k^\uparrow$ of
$\hat H$ upwards, the coherent
contribution $\WW(\hat\rho)-\WW(\DD\hat\rho)\geq0$ follows by majorization.

\begin{figure}[t]
\includegraphics[width=1.0\columnwidth]{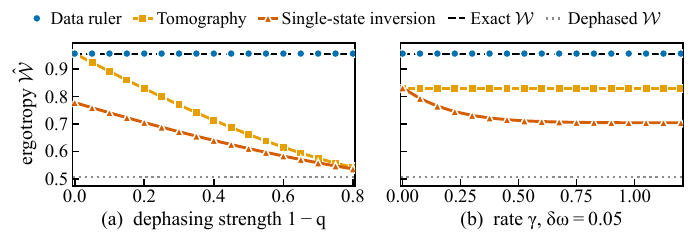}
\caption{Ergotropy from the ruler (circles), direct tomography at $t=0$
(squares), and single-state inversion with level error $\delta\omega=0.05$
(triangles), vs.\ (a) the common-mode dephasing strength $1-q$ ($q$ is the
retained coherence) and (b) the evolution dephasing
$\gamma$. Dashed: exact $\WW$; dotted: $\WW(\DD\hat\rho_0)$. Three levels,
$\hat H=\mathrm{diag}(0,1.37,3.04)$.\label{fig:coherence}}
\end{figure}

Figure~\ref{fig:coherence} quantifies this for three levels: the ruler
reproduces $\WW(\hat\rho_0)$ to machine precision, whereas tomography decays with the
preparation-and-readout dephasing and inversion is degraded further by the
calibration error; at dephasing strength $1-q=0.5$, $\gamma=0.8$,
$\delta\omega=0.05$ the coherent
contribution survives at $100.0\%$, $34.4\%$ and $22.9\%$, and the bias is
one-sided, so conventional estimates underestimate the advantage.
 On a concrete platform --- three transmons on one cavity, $70\%$
of the coherence surviving preparation --- tomography saturates at a one-sided
$-34\%$ error that no shot budget removes, whereas the ruler is unbiased to
$5\%$ after $\sim3\times10^5$ shots, about $30$ s (Sec.~S9 \cite{supp}).
\section{Conclusion}

We have introduced the quantum data ruler: a known reference whose measured
signal serves as a calibrated scale, against which the signal of an unknown
state is compared, so that a matrix element of the latter is read off from the
ratio. It is exact at arbitrary time and for arbitrary Hamiltonians, because the expectation-value change of an eigen-operator
factorizes into a state-dependent constant times a single time function. Every
pair of levels carries a level-pair qubit whose readout precision obeys
$\Delta \hat X_{mn}\Delta \hat Y_{mn}\geq|p_m-p_n|$, with the incompatibility of joint
estimation switched off at population balance (Theorem~2). The ratio is unbiased
under any noise whose jump operators are diagonal or rank-one in the energy
basis, needs no knowledge of the levels, and recovers the coherence that the
two-point measurement scheme erases; its classical image is the action--angle
Fourier mode, integrable and noiseless.

Equation~\eqref{eq:ratio} is symmetric in $\hat\rho_0$ and $\hat\rho_r$: a state read out
in one experiment serves as the ruler in the next, so every state with any
coherence --- all but $\hat\rho_0=\sum_n p_n|E_n\rangle\langle E_n|$ --- can be turned
into a data ruler, and quantum data enter a shared pool in which states calibrate
one another \cite{RevModPhys.79.555}.

\begin{acknowledgments}
This work was supported by NSFC (Grant No.~12205092), the Hunan
Provincial Natural Science Foundation of China (Grant
Nos.~2026JJ50100) and the Scientific Research Fund of
Hunan Provincial Education Department (Grant No.~24B0670).
\end{acknowledgments}

\bibliography{qdr}

@PREAMBLE{
 "\providecommand{\noopsort}[1]{}" 
 # "\providecommand{\singleletter}[1]{#1}%" 
}

@article{Bohr1928-BOHTQP,
	author = {Niels Bohr},
	journal = {Nature},
	pages = {580--590},
	title = {The Quantum Postulate and the Recent Development of Atomic Theory},
	volume = {121},
	year = {1928},
	doi={doi.org/10.1038/121580a0}
}

@book{von2018mathematical,
	title={Mathematical foundations of quantum mechanics: New edition},
	author={Von Neumann, John},
	year={2018},
	publisher={Princeton university press}
}

@article{RevModPhys.75.715,
	title = {Decoherence, einselection, and the quantum origins of the classical},
	author = {Zurek, Wojciech Hubert},
	journal = {Rev. Mod. Phys.},
	volume = {75},
	issue = {3},
	pages = {715--775},
	numpages = {0},
	year = {2003},
	month = {May},
	publisher = {American Physical Society},
	doi = {10.1103/RevModPhys.75.715},
	url = {https://link.aps.org/doi/10.1103/RevModPhys.75.715}
}

@book{schlosshauer2007decoherence,
	title={Decoherence and the quantum-to-classical transition},
	author={Schlosshauer, Maximilian},
	publisher={Springer}
}

@article{braginsky1980quantum,
	title={Quantum nondemolition measurements},
	author={Braginsky, Vladimir B and Vorontsov, Yuri I and Thorne, Kip S},
	journal={Science},
	volume={209},
	number={4456},
	pages={547--557},
	year={1980},
	publisher={American Association for the Advancement of Science}
}

@article{grangier1998quantum,
	title={Quantum non-demolition measurements in optics},
	author={Grangier, Philippe and Levenson, Juan Ariel and Poizat, Jean-Philippe},
	journal={Nature},
	volume={396},
	number={6711},
	pages={537--542},
	year={1998},
	publisher={Nature Publishing Group UK London},
	doi={doi.org/10.1038/25059}
}

@book{leonhardt1997measuring,
	title={Measuring the quantum state of light},
	author={Leonhardt, Ulf},
	volume={22},
	year={1997},
	publisher={Cambridge university press}
}

@article{RevModPhys.81.299,
	title = {Continuous-variable optical quantum-state tomography},
	author = {Lvovsky, A. I. and Raymer, M. G.},
	journal = {Rev. Mod. Phys.},
	volume = {81},
	issue = {1},
	pages = {299--332},
	numpages = {0},
	year = {2009},
	month = {Mar},
	publisher = {American Physical Society},
	doi = {10.1103/RevModPhys.81.299},
	url = {https://link.aps.org/doi/10.1103/RevModPhys.81.299}
}

@article{PhysRevLett.60.1351,
	title = {How the result of a measurement of a component of the spin of a spin-1/2 particle can turn out to be 100},
	author = {Aharonov, Yakir and Albert, David Z. and Vaidman, Lev},
	journal = {Phys. Rev. Lett.},
	volume = {60},
	issue = {14},
	pages = {1351--1354},
	numpages = {0},
	year = {1988},
	month = {Apr},
	publisher = {American Physical Society},
	doi = {10.1103/PhysRevLett.60.1351},
	url = {https://link.aps.org/doi/10.1103/PhysRevLett.60.1351}
}

@article{PhysRevLett.66.1107,
	title = {Realization of a measurement of a ``weak value''},
	author = {Ritchie, N. W. M. and Story, J. G. and Hulet, Randall G.},
	journal = {Phys. Rev. Lett.},
	volume = {66},
	issue = {9},
	pages = {1107--1110},
	numpages = {0},
	year = {1991},
	month = {Mar},
	publisher = {American Physical Society},
	doi = {10.1103/PhysRevLett.66.1107},
	url = {https://link.aps.org/doi/10.1103/PhysRevLett.66.1107}
}

@article{lundeen2011direct,
	title={Direct measurement of the quantum wavefunction},
	author={Lundeen, Jeff S and Sutherland, Brandon and Patel, Aabid and Stewart, Corey and Bamber, Charles},
	journal={Nature},
	volume={474},
	number={7350},
	pages={188--191},
	year={2011},
	publisher={Nature Publishing Group UK London},
	doi={doi.org/10.1038/nature10120}
}

@article{PhysRevLett.108.070402,
	title = {Procedure for Direct Measurement of General Quantum States Using Weak Measurement},
	author = {Lundeen, Jeff S. and Bamber, Charles},
	journal = {Phys. Rev. Lett.},
	volume = {108},
	issue = {7},
	pages = {070402},
	numpages = {5},
	year = {2012},
	month = {Feb},
	publisher = {American Physical Society},
	doi = {10.1103/PhysRevLett.108.070402},
	url = {https://link.aps.org/doi/10.1103/PhysRevLett.108.070402}
}

@article{jordan2020quantum,
	title={Quantum measurement engines and their relevance for quantum interpretations},
	author={Jordan, Andrew N and Elouard, Cyril and Auff{\`e}ves, Alexia},
	journal={Quantum Studies: Mathematics and Foundations},
	volume={7},
	number={2},
	pages={203--215},
	year={2020},
	publisher={Springer},
	doi={doi.org/10.1007/s40509-019-00217-2}
}

@article{weinacht1999controlling,
	title={Controlling the shape of a quantum wavefunction},
	author={Weinacht, TC and Ahn, Jaewook and Bucksbaum, Phil H},
	journal={Nature},
	volume={397},
	number={6716},
	pages={233--235},
	year={1999},
	publisher={Nature Publishing Group UK London},
	doi={doi.org/10.1038/16654}
}

@article{PhysRevLett.80.1418,
	title = {Quantum State Holography},
	author = {Leichtle, C. and Schleich, W. P. and Averbukh, I. Sh. and Shapiro, M.},
	journal = {Phys. Rev. Lett.},
	volume = {80},
	issue = {7},
	pages = {1418--1421},
	numpages = {0},
	year = {1998},
	month = {Feb},
	publisher = {American Physical Society},
	doi = {10.1103/PhysRevLett.80.1418},
	url = {https://link.aps.org/doi/10.1103/PhysRevLett.80.1418}
}

@article{Lu2022InstMeas,
  title = {Instantaneous indirect measurement principle in quantum mechanics},
  author = {Lu, Wangjun and Zhang, Xingyu and Shao, Lei and Zhang, Zhucheng and
            Chen, Jie and Zhang, Rui and Xiong, Shaojie and Zhan, Liyao and
            Wang, Xiaoguang},
  journal = {arXiv:2207.04761}, year = {2022}
}

@article{PhysRevA.104.043706,
	title = {Optimal state for a Tavis-Cummings quantum battery via the Bethe ansatz method},
	author = {Lu, Wangjun and Chen, Jie and Kuang, Le-Man and Wang, Xiaoguang},
	journal = {Phys. Rev. A},
	volume = {104},
	issue = {4},
	pages = {043706},
	numpages = {14},
	year = {2021},
	month = {Oct},
	publisher = {American Physical Society},
	doi = {10.1103/PhysRevA.104.043706},
	url = {https://link.aps.org/doi/10.1103/PhysRevA.104.043706}
}

@article{PhysRevA.53.2038,
	title = {Quantum-state disturbance versus information gain: Uncertainty relations for quantum information},
	author = {Fuchs, Christopher A. and Peres, Asher},
	journal = {Phys. Rev. A},
	volume = {53},
	issue = {4},
	pages = {2038--2045},
	year = {1996},
	month = {Apr},
	publisher = {American Physical Society},
	doi = {10.1103/PhysRevA.53.2038},
	url = {https://link.aps.org/doi/10.1103/PhysRevA.53.2038}
}

@article{PhysRevLett.86.1366,
	title = {Fidelity balance in quantum operations},
	author = {Banaszek, Konrad},
	journal = {Phys. Rev. Lett.},
	volume = {86},
	issue = {7},
	pages = {1366--1369},
	year = {2001},
	month = {Feb},
	publisher = {American Physical Society},
	doi = {10.1103/PhysRevLett.86.1366},
	url = {https://link.aps.org/doi/10.1103/PhysRevLett.86.1366}
}

@article{PhysRevLett.100.210504,
	title = {Global information balance in quantum measurements},
	author = {Buscemi, Francesco and Hayashi, Masahito and Horodecki, Micha{\l}},
	journal = {Phys. Rev. Lett.},
	volume = {100},
	issue = {21},
	pages = {210504},
	year = {2008},
	month = {May},
	publisher = {American Physical Society},
	doi = {10.1103/PhysRevLett.100.210504},
	url = {https://link.aps.org/doi/10.1103/PhysRevLett.100.210504}
}

@article{RevModPhys.79.555,
	title = {Reference frames, superselection rules, and quantum information},
	author = {Bartlett, Stephen D. and Rudolph, Terry and Spekkens, Robert W.},
	journal = {Rev. Mod. Phys.},
	volume = {79},
	issue = {2},
	pages = {555--609},
	year = {2007},
	month = {Apr},
	publisher = {American Physical Society},
	doi = {10.1103/RevModPhys.79.555},
	url = {https://link.aps.org/doi/10.1103/RevModPhys.79.555}
}

@article{PerarnauLlobet2017,
  title = {No-Go Theorem for the Characterization of Work Fluctuations in Coherent Quantum Systems},
  author = {Perarnau-Llobet, Mart{\i} and B{\"a}umer, Elisa and Hovhannisyan, Karen V. and Huber, Marcus and Acin, Antonio},
  journal = {Phys. Rev. Lett.},
  volume = {118}, issue = {7}, pages = {070601},
  year = {2017}, doi = {10.1103/PhysRevLett.118.070601}
}

@article{Streltsov2017,
  title = {Quantum coherence as a resource},
  author = {Streltsov, Alexander and Adesso, Gerardo and Plenio, Martin B.},
  journal = {Rev. Mod. Phys.},
  volume = {89}, issue = {4}, pages = {041003},
  year = {2017}, doi = {10.1103/RevModPhys.89.041003}
}

@article{Elouard2026,
  title = {Minimally invasive measurement of work in coherent quantum systems},
  author = {Elouard, Cyril and Hovhannisyan, Karen and Rubino, Giulia},
  journal = {arXiv:2607.02652}, year = {2026}
}

@article{Aguilar2026,
  title = {Minimal-backaction work statistics of coherent engines},
  author = {Aguilar, Milton and Rodrigues, Franklin L. S. and Lutz, Eric},
  journal = {arXiv:2603.01962}, year = {2026}
}

@article{Damas2026,
  title = {Shortcut-error signatures in coherence-retaining endpoint work quasistatistics},
  author = {Damas, Gabriella G. and de Moraes Neto, G. D.},
  journal = {Phys. Lett. A}, year = {2026}, doi = {10.1016/j.physleta.2026.132220}
}

@article{Zhang2026QB,
  title = {Quantum batteries in coherent Ising machine},
  author = {Zhang, Jin-Tian and Ma, Shuang-Quan and Jin, Jing-Yi-Ran and Liu, Tao and Ai, Qing},
  journal = {Phys. Rev. Research}, volume = {8}, issue = {2}, pages = {023187},
  year = {2026}, doi = {10.1103/4bj4-4fpy}
}

@article{Wang2026Multiparam,
  title = {Minimal Trade-off and Optimal Measurement for Multiparameter Quantum Estimation},
  author = {Wang, Lingna and Chen, Hongzhen and Yuan, Haidong},
  journal = {Phys. Rev. Lett.}, volume = {137}, issue = {2}, pages = {020804},
  year = {2026}, doi = {10.1103/dvj7-9thf}
}

@article{Chen2025Probe,
  title = {Quantum Probe Tomography},
  author = {Chen, Sitan and Cotler, Jordan and Huang, Hsin-Yuan},
  journal = {arXiv:2510.08499}, year = {2025}
}

@unpublished{supp,
  title = {{See Supplemental Material at [URL will be inserted by publisher] for the proofs of Theorems 1, 2, and 3, the sample-complexity analysis, the comparison with established readout schemes, the classical correspondence, and platform-specific numerical studies}},
  year = {2026},
}

@book{FultonHarris1991,
  title = {Representation Theory: A First Course},
  author = {Fulton, William and Harris, Joe},
  publisher = {Springer}, series = {Graduate Texts in Mathematics}, volume = {129},
  year = {1991}
}

@article{Feynman1957,
  title = {Geometrical Representation of the Schr{\"o}dinger Equation for Solving Maser Problems},
  author = {Feynman, R. P. and Vernon, F. L. and Hellwarth, R. W.},
  journal = {J. Appl. Phys.}, volume = {28}, issue = {1}, pages = {49},
  year = {1957}, doi = {10.1063/1.1722572}
}

@article{Chrysosthemos2018,
  title = {Quantum coherence--classical uncertainty tradeoff relations},
  author = {Chrysosthemos, Diego S. S. and Maziero, Jonas},
  journal = {arXiv:1809.06354}, year = {2018}
}

@article{Kurzynski2011,
  title = {Multi-Bloch vector representation of the qutrit},
  author = {Kurzy{\'n}ski, Pawe{\l}},
  journal = {Quantum Inf. Comput.}, volume = {11}, issue = {5-6}, pages = {361},
  year = {2011}, note = {arXiv:0912.3155}
}

@article{LopezSaldivar2019,
  title = {Qubit representation of qudit states: correlations and state reconstruction},
  author = {L{\'o}pez-Sald{\'i}var, Julio A. and Casta{\~n}os, Octavio and Man'ko, Margarita A. and Man'ko, Vladimir I.},
  journal = {Quantum Inf. Process.}, volume = {18}, issue = {7}, pages = {210},
  year = {2019}, doi = {10.1007/s11128-019-2327-1}, note = {arXiv:1905.04905}
}

@article{Brunner2012,
  title = {Virtual qubits, virtual temperatures, and the foundations of thermodynamics},
  author = {Brunner, Nicolas and Linden, Noah and Popescu, Sandu and Skrzypczyk, Paul},
  journal = {Phys. Rev. E}, volume = {85}, issue = {5}, pages = {051117},
  year = {2012}, doi = {10.1103/PhysRevE.85.051117}, note = {arXiv:1106.2138}
}

@article{Robertson1929,
  title = {The uncertainty principle},
  author = {Robertson, H. P.},
  journal = {Phys. Rev.}, volume = {34}, issue = {1}, pages = {163},
  year = {1929}, doi = {10.1103/PhysRev.34.163}
}

@article{Suzuki2016,
  title = {Explicit formula for the Holevo bound for two-parameter qubit-state estimation},
  author = {Suzuki, Jun},
  journal = {J. Phys. A: Math. Theor.}, volume = {49}, issue = {44}, pages = {445302},
  year = {2016}, doi = {10.1088/1751-8113/49/44/445302}
}

@article{Ragy2016,
  title = {Compatibility in multiparameter quantum metrology: The Holevo bound and beyond},
  author = {Ragy, Sammy and Jarzyna, Marcin and Demkowicz-Dobrza{\'n}ski, Rafa{\l}},
  journal = {Phys. Rev. A}, volume = {94}, issue = {5}, pages = {052108},
  year = {2016}, doi = {10.1103/PhysRevA.94.052108}
}

@book{Holevo2011,
  title = {Probabilistic and Statistical Aspects of Quantum Theory},
  author = {Holevo, Alexander S.}, publisher = {Springer}, series = {Publications of the Scuola Normale Superiore},
  year = {2011}, doi = {10.1007/978-88-7642-378-9}
}

@article{Szczykulska2016,
  title = {Multi-parameter quantum metrology},
  author = {Szczykulska, Magdalena and Baumgratz, Tillmann and Datta, Animesh},
  journal = {Adv. Phys. X}, volume = {1}, issue = {4}, pages = {621},
  year = {2016}, doi = {10.1080/23746149.2016.1230476}
}

@article{Blais2021,
  title = {Circuit quantum electrodynamics},
  author = {Blais, Alexandre and Grimsmo, Arne L. and Girvin, S. M. and Wallraff, Andreas},
  journal = {Rev. Mod. Phys.}, volume = {93}, issue = {2}, pages = {025005},
  year = {2021}, doi = {10.1103/RevModPhys.93.025005}
}

@article{Paik2011,
  title = {Observation of High Coherence in Josephson Junction Qubits Measured in a Three-Dimensional Circuit {QED} Architecture},
  author = {Paik, Hanhee and Schuster, D. I. and Bishop, Lev S. and Kirchmair, Gerhard and Catelani, Gianluigi and Sears, A. P. and Johnson, B. R. and Reagor, M. J. and Frunzio, L. and Glazman, L. I. and Girvin, S. M. and Schoelkopf, R. J.},
  journal = {Phys. Rev. Lett.}, volume = {107}, issue = {24}, pages = {240501},
  year = {2011}, doi = {10.1103/PhysRevLett.107.240501}
}

@book{Breuer2002,
  title = {The Theory of Open Quantum Systems},
  author = {Breuer, Heinz-Peter and Petruccione, Francesco},
  publisher = {Oxford University Press}, year = {2002}
}

\end{document}


\title{Supplemental Material for ``Quantum Data Ruler''}
\author{Wangjun Lu}
\date{\today}
\begin{center}
{\large\bfseries Supplemental Material for ``Quantum Data Ruler''}\\[5pt]
Wangjun Lu\\[2pt]
(\today)
\end{center}
\vspace{4pt}

\setcounter{equation}{0}
\renewcommand{\theequation}{S\arabic{equation}}
\setcounter{figure}{0}
\renewcommand{\thefigure}{S\arabic{figure}}
\setcounter{table}{0}
\renewcommand{\thetable}{S\arabic{table}}

This Supplemental Material contains: (S1) the proof of Theorem~3 on
dephasing-unbiasedness, together with a precise statement of what the theorem
does and does not remove; (S2) the sample complexity of the readout and its
statistical error; (S3) a systematic comparison with established readout
schemes; (S4) the application to coherent work and coherent ergotropy, including
the majorization argument and the numerical comparison behind Fig.~2 of the main
text; (S5) the application to calibration-free spectral diagnosis and its
relation to echo and dynamical decoupling; (S6) the universality numerics and
the information--disturbance trade-off; (S7) the classical limit; (S8) the
feasibility hierarchy of the readout and further directions; (S9) a concrete
experimental proposal for a Tavis--Cummings quantum battery on circuit-QED
hardware, with parameters, shot budget and error budget; (S10) the proof of
Theorem~2 on the joint-estimation incompatibility, with the single-copy cost;
(S11) the general noise-immunity criterion behind Theorem~3, its channel-by-channel
test, the bound on the residual bias, and the platform numerics.
Sections are ordered by topic, not by the number of the theorem they support:
S1 supports Theorem~3 and S10 supports Theorem~2.

\makeatletter
\renewcommand{\@seccntformat}[1]{}
\makeatother
\renewcommand{\thesection}{}
\renewcommand{\thesubsection}{}

\setcounter{tocdepth}{2}
\tableofcontents

\section{S1. Dephasing-unbiasedness: proof and scope}

\subsection{S1.1 Setup}

Throughout, hats mark operators; a symbol without a hat is a complex number or a
classical quantity. Let $\hat H$ be Hermitian with eigenbasis
$\{|E_n\rangle\}$, $\hat H|E_n\rangle=E_n|E_n\rangle$, and let
\begin{equation}
\hat A_{mn}=|E_m\rangle\langle E_n|,\qquad \omega_{mn}=E_m-E_n,\qquad
p_n=\langle E_n|\hat\rho_0|E_n\rangle ,
\end{equation}
so that $\hat A_{mn}$ is the transition operator of the pair $(m,n)$,
$\omega_{mn}$ the transition frequency, and $p_n$ the initial population of
level $n$, whence $p_m-p_n$ is the population imbalance of that pair. For any
state $\hat\rho$ we write its energy-basis matrix elements as
$\rho_{nm}=\langle E_n|\hat\rho|E_m\rangle$; for the unknown initial state this
is $\rho_{nm}=\langle E_n|\hat\rho_0|E_m\rangle$ and for the known reference
$\rho_{r,nm}=\langle E_n|\hat\rho_r|E_m\rangle$. We assume that between
preparation and readout the off-diagonal elements in the energy eigenbasis decay
by a factor that does not depend on the state:
\begin{equation}
\rho_{nm}(t)=\rho_{nm}(0)\,e^{-i\omega_{mn}t}\,\phi_{mn}(t),\qquad m\neq n,
\label{eq:decay}
\end{equation}
where $\rho_{nm}(t)=\langle E_n|\hat\rho(t)|E_m\rangle$ refers to the evolved
unknown state $\hat\rho(t)$, and where $|\phi_{mn}(t)|\leq1$,
$\phi_{mn}(0)=1$, and $\phi_{mn}$ is the same
function for the unknown state $\hat\rho_0$ and for the reference state $\hat\rho_r$.
Equation~\eqref{eq:decay} covers
\begin{itemize}
\item a discrete phase-damping channel $\Phi_p$ applied once:
$\phi_{mn}=1-p$;
\item continuous Lindblad dephasing with rate $\gamma$ in the energy basis:
$\phi_{mn}(t)=e^{-\gamma t}$;
\item any inhomogeneous dephasing, i.e.\ a rate $\gamma_{mn}$ depending on the
pair: $\phi_{mn}(t)=e^{-\gamma_{mn}t}$.
\end{itemize}

\subsection{S1.2 Proof of Theorem 3}

Because $\hat A_{mn}=|E_m\rangle\langle E_n|$, the trace
$\Tr[\hat A_{mn}\hat\rho(t)]$ selects the matrix element
$\langle E_n|\hat\rho(t)|E_m\rangle$, denoted $\rho_{nm}(t)$ below.
Using Eq.~\eqref{eq:decay} for the unknown state,
\begin{equation}
\Davg{\hat A_{mn}}(t,\hat\rho_0)=\rho_{nm}(t)-\rho_{nm}(0)
=\rho_{nm}(0)\big[e^{-i\omega_{mn}t}\phi_{mn}(t)-1\big],
\end{equation}
and identically for the reference,
\begin{equation}
\Davg{\hat A_{mn}}(t,\hat\rho_r)=\rho_{r,nm}(0)\big[e^{-i\omega_{mn}t}\phi_{mn}(t)-1\big].
\end{equation}
The bracket is verbatim the same in both lines, so it cancels exactly in
the ratio:
\begin{equation}
\frac{\Davg{\hat A_{mn}}(t,\hat\rho_0)}{\Davg{\hat A_{mn}}(t,\hat\rho_r)}
=\frac{\rho_{nm}}{\rho_{r,nm}},
\qquad \forall\{\phi_{mn}\},\ \forall t .
\end{equation}
Multiplying by the known $\rho_{r,nm}$ gives $\rho_{nm}$ with no reference to
$\omega_{mn}$, to $t$, or to the dephasing.

Two structural remarks. (i) The cancellation works because the factor is common
to the same pair $(m,n)$; homogeneity over pairs is not needed. What
is needed is that the dephasing be diagonal-projecting (it must not mix
off-diagonal elements among themselves) and common-mode (identical for $\hat\rho_0$
and $\hat\rho_r$). (ii) Equivalently, the dephasing superoperator must commute with
the evolution $e^{-i\hat Ht}\cdot e^{i\hat Ht}$; this is what makes the problem a single
common scalar factor per pair rather than a matrix.

\subsection{S1.3 What is unbiased and what is not}

It is essential to separate three different quantities, all of which one might
loosely call ``the spectrum'':

\begin{table}[h]
\caption{Three layers of ``spectral unbiasedness'' under dephasing.}
\begin{ruledtabular}
\begin{tabular}{lcc}
Quantity & Unbiased under dephasing? & Reason \\
\hline
Peak position $\omega_{mn}$ & Yes & dephasing multiplies the amplitude, \\
 & & it does not shift the oscillation frequency \\
Peak amplitude $\rho_{nm}$ & Yes, after the ratio calibration & the common factor \\
 & & $\phi_{mn}(t)$ cancels in the ratio \\
Peak width (resolution) & No & dephasing broadens a $\delta$ peak \\
 & & into a Lorentzian of width $\sim\gamma$\\
\end{tabular}
\end{ruledtabular}
\end{table}

The last line is a genuine loss, not a bias: no rescaling restores the linewidth.
The ruler converts a systematic error into no error, but it cannot convert
lost information into recovered information.

\section{S2. Sample complexity and statistical error}

\paragraph{Measurement settings.} Reading one element $\rho_{nm}$ requires the two
Hermitian readouts $\hat X_{mn}$ and $\hat Y_{mn}$, i.e.\ $O(1)$ settings. Reading all
$d(d-1)/2$ independent off-diagonal elements requires $O(d^2)$ settings, the
same order as full tomography.

\paragraph{Time sampling and repetitions.} Resolving a single frequency $\omega_{mn}$
to resolution $\Delta\omega$ needs $N_t=O(\omega_{\max}/\Delta\omega)$ time
points; the shot-noise-limited precision $\epsilon$ at total interrogation time
$T$ obeys the Cram\'er--Rao scaling $\Delta\omega\gtrsim1/(T\sqrt{N_{\rm rep}})$,
so $N_{\rm rep}=O(\epsilon^{-2})$. The full spectrum costs $O(d^2/\epsilon^2)$
samples.

\paragraph{Honest assessment.} The ruler does not beat tomography in sample
complexity, and it does not address the exponential cost of characterizing a
many-body state; randomized-measurement and classical-shadow protocols
\cite{leonhardt1997measuring,RevModPhys.81.299} are the right tools there. The
ruler's advantages are of a different kind: (i) it needs no knowledge of the
energy levels of $\hat H$, because the reference state is the only calibration;
(ii) it is unbiased under dephasing of unknown strength, without echo or
dynamical decoupling; (iii) it returns the phase of $\rho_{nm}$, which
scalar coherence measures do not; (iv) it needs no post-selection.

\section{S3. Comparison with established readout schemes}

Table~\ref{tab:comparison} places the ruler among the standard schemes.
``Coherence'' means reading the off-diagonal element $\rho_{nm}$;
``dephasing-unbiased'' is assessed for that readout, which dephasing would
otherwise bias.

\begin{table*}[t]
\caption{Comparison of the data ruler with established readout schemes.
$\checkmark$: supported; $\times$: not supported; $-$: not applicable.}
\label{tab:comparison}
\begin{ruledtabular}
\begin{tabular}{lccccc}
Scheme & Coherence & Any $t$ & No levels & Deph.-unbiased & Mixed\\
\hline
QND & $\times$ & $\checkmark$ & $\times$ & $-$ & $\checkmark$\\
Weak measurement & $\checkmark$ & $\times$ & $-$ & $\times$ & $\times$\\
State holography & $\checkmark$ & $\checkmark$ & $\times$ & $\times$ & $\times$\\
Tomography & $\checkmark$ & $\checkmark$ & $\times$ & $\times$ & $\checkmark$\\
Classical shadows & $\checkmark$ & $\checkmark$ & $\times$ & $\times$ & $\checkmark$\\
KD / MH quasiprobabilities & $\checkmark$ & $\checkmark$ & $\times$ & $\times$ & $\checkmark$\\
Data ruler (this work) & $\checkmark$ & $\checkmark$ & $\checkmark$ & $\checkmark$ & $\checkmark$\\
\end{tabular}
\end{ruledtabular}
\end{table*}

Two comparisons deserve a comment. Against weak measurement, the ruler
needs neither a weak-coupling limit nor post-selection: the coupling strength and
the interrogation time drop out of the ratio, so there is no trade-off between
signal strength and back-action, and no post-selection loss. Against
KD/MH quasiprobability reconstructions of work statistics, which also
preserve coherence sensitivity \cite{Damas2026}, the ruler is complementary:
quasiprobabilities reconstruct a full (possibly non-positive) distribution,
whereas the ruler returns the specific complex numbers $\rho_{nm}$ that enter
energetic quantities, and does so without knowing the levels of $\hat H$.

\section{S4. Application I: coherent work and coherent ergotropy}

\subsection{S4.1 The problem}

In the two-point measurement (TPM) scheme the first projective measurement of
$\hat H(0)$ replaces $\hat\rho_0$ by its energy-dephased counterpart $\DD(\hat\rho_0)$. For a
closed driven system the mean work is
$W=\int dt\,\operatorname{Tr}[\dot{\hat H}(t)\hat\rho(t)]$, so the TPM value differs from
the unmeasured one by
\begin{equation}
W_{\rm coh}=\int dt\,\operatorname{Tr}\big[\dot{\hat H}(t)\big(\hat\rho(t)-\DD(\hat\rho(t))\big)\big],
\end{equation}
the entire coherence contribution. This erasure is not a technical artifact: no
universal positive work distribution can reproduce TPM on incoherent states
while giving the correct mean for arbitrary coherent ones
\cite{PerarnauLlobet2017}. Operationally, the back-action can be strong enough
to change the mode of operation of a coherent machine, which may cease to work
as an engine under observation \cite{Aguilar2026}. Recent proposals mitigate this
by restructuring the measurement itself, e.g.\ by using the Heisenberg-picture
Hamiltonian variation \cite{Elouard2026} or a dynamic-Bayesian-network update
\cite{Aguilar2026}; both recover coherent work statistics while keeping
positivity.

The ruler offers a different mechanism. It does not reconstruct a work
distribution at all; it reads the complex numbers $\rho_{nm}$ on which
$W_{\rm coh}$ depends, without projecting $\hat\rho_0$ and without knowing the levels
of $\hat H$. Its specific advantage is that the reading stays unbiased when the
dephasing strength is unknown (Theorem~3), which is the regime in which
conventional readouts acquire a one-sided systematic error.

\subsection{S4.2 Coherent ergotropy and majorization}

For a battery with Hamiltonian $\hat H=\sum_n E_n|E_n\rangle\langle E_n|$, the
ergotropy is
$\WW(\hat\rho_0)=\operatorname{Tr}(\hat\rho_0\hat H)-\sum_k\lambda_k^\downarrow E_k^\uparrow$,
where $\{\lambda_k^\downarrow\}$ are the eigenvalues of $\hat\rho_0$ in decreasing
order and $\{E_k^\uparrow\}$ the energies in increasing order. Since $\hat H$ is
diagonal, $\operatorname{Tr}(\hat\rho_0\hat H)=\operatorname{Tr}(\DD(\hat\rho_0)\,\hat H)$, so
\begin{equation}
\WW(\hat\rho_0)-\WW(\DD(\hat\rho_0))=\sum_k\mu_k^\downarrow E_k^\uparrow
-\sum_k\lambda_k^\downarrow E_k^\uparrow ,
\end{equation}
with $\{\mu_k\}$ the eigenvalues of $\DD(\hat\rho_0)$, i.e.\ the diagonal of $\hat\rho_0$ in
the energy basis. By the Schur--Horn theorem the eigenvalue vector
$\lambda$ majorizes the diagonal $\mu$; since $E^\uparrow$ is increasing,
$\lambda\succ\mu$ implies $\sum_k\lambda_k^\downarrow E_k^\uparrow\leq
\sum_k\mu_k^\downarrow E_k^\uparrow$. Hence
\begin{equation}
\WW(\hat\rho_0)\geq\WW(\DD(\hat\rho_0))
\end{equation}
always: coherence never reduces the maximum extractable work, and the difference
is exactly the coherent ergotropy. This is the quantity that recent work has
identified as the slow-decaying, robust part of a quantum battery's stored
energy \cite{Zhang2026QB}, and which experiments currently access only by
destroying coherence in a control run \cite{Streltsov2017}.

\subsection{S4.3 Numerical comparison}

We take a three-level system with $\hat H=\mathrm{diag}(0,1.37,3.04)$ (non-equidistant,
non-resonant) and a randomly drawn mixed $\hat\rho_0$, for which
$\WW(\hat\rho_0)=0.957174$, $\WW(\DD(\hat\rho_0))=0.507759$, and the coherent part is
$0.449414$, i.e.\ $47.0\%$ of the ergotropy. The readout is performed at
$t=\pi/\omega_{10}=2.2931$, where the signal is of order unity. Three schemes are
compared under a common-mode dephasing that retains a fraction $q$ of the
coherences between preparation and readout (so $q=1$ is ideal and the dephasing
strength is $1-q$), an unknown evolution dephasing rate $\gamma$, and a
level-calibration
error $\delta\omega$ in the frequency used to invert a single-state dynamical
signal:

\begin{itemize}
\item \textbf{Data ruler}: $\rho_{nm}^{\rm est}=\rho_{r,nm}\Davg{\hat A_{mn}}(t,\hat\rho_0)/\Davg{\hat A_{mn}}(t,\hat\rho_r)$;
\item \textbf{Direct tomography}: measure $\langle \hat X_{mn}\rangle,\langle \hat Y_{mn}\rangle$ at $t=0$;
\item \textbf{Single-state inversion}: $\rho_{nm}^{\rm est}=\Davg{\hat A_{mn}}(t,\hat\rho_0)/(e^{i\omega_{mn}^{\rm est}t}-1)$,
which requires the levels and assumes no dephasing.
\end{itemize}

\begin{table}[h]
\caption{Recovered ergotropy $\WW^{\rm est}$, with the surviving fraction of the
coherent contribution, $(\WW^{\rm est}-\WW(\DD(\hat\rho_0)))/(\WW(\hat\rho_0)-\WW(\DD(\hat\rho_0)))$, in
parentheses, versus the common-mode dephasing strength $1-q$ ($q$ is the
retained coherence). ``Ruler'' is the data ruler, ``Direct'' is tomography at
$t=0$, and
``Inversion'' is single-state dynamical inversion.}
\begin{ruledtabular}
\small
\begin{tabular}{cccccc}
$1-q$ & $\gamma$ & $\delta\omega$ & Ruler & Direct & Inversion\\
\hline
0.00 & 0.00 & 0.00 & 0.9572 (100.0\%) & 0.9572 (100.0\%) & 0.9572 (100.0\%)\\
0.00 & 0.80 & 0.05 & 0.9572 (100.0\%) & 0.9572 (100.0\%) & 0.7783 (60.2\%)\\
0.20 & 0.80 & 0.05 & 0.9572 (100.0\%) & 0.8297 (71.6\%) & 0.7057 (44.0\%)\\
0.50 & 0.80 & 0.05 & 0.9572 (100.0\%) & 0.6622 (34.4\%) & 0.6107 (22.9\%)\\
0.50 & 0.20 & 0.05 & 0.9572 (100.0\%) & 0.6622 (34.4\%) & 0.6250 (26.1\%)\\
\end{tabular}
\end{ruledtabular}
\end{table}

The recovered $\rho_{nm}$ agrees with the exact value to $10^{-16}$ for the
ruler in every row, whereas direct tomography degrades linearly in the
dephasing strength $1-q$ and
single-state inversion degrades in both $1-q$ and $\gamma$ on top of the
calibration error. The bias is one-sided: every conventional estimate lies
below $\WW(\hat\rho_0)$, so the quantum advantage is systematically under-reported.
Figure~2 of the main text shows the same data as continuous scans.

A comment on scope: the ruler reads initial coherences $\rho_{nm}$ in the
eigenbasis of $\hat H(0)$. For a driven system the full $W_{\rm coh}$ involves the
coherence throughout the drive; recovering it requires either repeating the
readout at several times or, when the drive is known, propagating the measured
initial data. The ruler is thus the input stage of a coherent-work protocol, not
a replacement for the entire protocol.

\section{S5. Application II: calibration-free spectral diagnosis}

Because Eq.~(2) of the main text is independent of $\omega_{mn}$, the ruler reads
$\rho_{nm}$ without ever measuring the level spacing, and Theorem~3 keeps this
reading unbiased when the dephasing rate is unknown. This suggests a spectral
diagnostic that is complementary to the standard tools:

\begin{itemize}
\item \textbf{Echo and dynamical decoupling} (CPMG and relatives) suppress
dephasing by reversing it in the time domain; they require pulse sequences and
a known or inferable noise spectrum.
\item \textbf{The ruler} cancels the dephasing by taking a ratio between two
states that share the same bath. It requires no pulses, but it does require
preparing a known reference and it assumes common-mode noise.
\end{itemize}

The regime in which the ruler is preferable is therefore the one in which an
echo is unavailable or too costly: continuous-readout platforms, systems whose
dephasing cannot be reversed by the available controls, and situations where the
level structure is not calibrated in the first place. Conversely, when an echo
is cheap and the levels are known, echo-based protocols remain preferable; the
two tools are complementary rather than competing. We stress that the
common-mode assumption is the strongest hypothesis of the present scheme and
must be checked experimentally, e.g.\ by interleaving the unknown and reference
runs in time so that both sample the same noise realization.

For many-body spectral functions (dynamical structure factors), the ruler is
not directly applicable: it would require reading every transition
operator $|E_m\rangle\langle E_n|$ in an exponentially large space. It becomes
relevant in the local-probe setting, where a small, well-controlled probe
couples to a large system and the effective dimension stays small
\cite{Chen2025Probe}; there the freedom from calibration and from common-mode
noise is precisely what is needed.

\section{S6. Universality and the trade-off in detail}

\paragraph{On the status of the level-pair algebra.}
Observation~1 of the main text is stated with its status made explicit there,
and we repeat it here for referees. The fact that
$\hat X_{mn}=|E_m\rangle\langle E_n|+|E_n\rangle\langle E_m|$,
$\hat Y_{mn}=i(|E_n\rangle\langle E_m|-|E_m\rangle\langle E_n|)$ and
$\hat Z_{mn}=|E_m\rangle\langle E_m|-|E_n\rangle\langle E_n|$ obey the Pauli algebra
for every pair of levels is standard and is not claimed as a result of
this work. In the language of Lie theory it is the $\mathfrak{su}(2)$ triple
attached to a pair of roots, $(E_\alpha,E_{-\alpha},H_\alpha=[E_\alpha,E_{-\alpha}])$,
which is part of the Chevalley--Serre presentation of any semisimple Lie algebra
\cite{FultonHarris1991}; in explicit $d\times d$ matrices these are the
off-diagonal members of the Gell-Mann basis; for a genuine two-level system the
same mapping back to a spin goes back to Feynman, Vernon and Hellwarth
\cite{Feynman1957}. Reading $\rho_{nm}$ as $(\langle \hat X_{mn}\rangle+i\langle
\hat Y_{mn}\rangle)/2$ is likewise the ordinary business of Pauli-basis (Gell-Mann
basis) tomography, and is how coherence functions are accessed experimentally,
e.g.\ by Ramsey-type sequences in cold-atom systems. Even the Robertson bound
$\Delta \hat X_{mn}\Delta \hat Y_{mn}\geq|p_m-p_n|$ sits inside the well-studied family of
coherence--population trade-off relations \cite{Chrysosthemos2018}.

\paragraph{Relation to qubit decompositions of qudits.} We should also point out
that the framing --- describing a $d$-level system by one qubit, i.e.\ one
Bloch vector, per pair of levels --- has been developed before, and for the same
purpose of state reconstruction. Kurzy\'nski represented a qutrit by three
qubit-like Bloch vectors \cite{Kurzynski2011}, and L\'opez-Sald\'ivar,
Casta\~nos and Man'ko generalised this to a decomposition of an arbitrary qudit
into $d(d-1)/2$ qubits, used the positivity conditions to derive inequalities
obeyed by the density-matrix components, and applied them to state
reconstruction \cite{LopezSaldivar2019}. The word virtual qubit that one
might be tempted to use for our level pair is already standard in quantum
thermodynamics with a different meaning, namely a two-level subsystem of the
composite of two thermal baths \cite{Brunner2012}; we therefore avoid it.

Two differences keep our use of the structure distinct from these works. (i) The
inequalities of Ref.~\cite{LopezSaldivar2019} are positivity constraints:
they bound the matrix elements a physical state may have, the prototype being
$|\rho_{nm}|^2\leq p_mp_n$. Our Eq.~(5) is of a different kind: it is a
precision bound on the simultaneous readout of the two quadratures of
$\rho_{nm}$ on an arbitrary state, physical or not, and it follows from the
non-commutativity of $\hat X_{mn}$ and $\hat Y_{mn}$. The two are frequently confused and
should not be: positivity says how large a coherence may be, the Robertson bound
says how precisely one can read a coherence that is there. (ii) Neither work
considers a dynamical, reference-based readout: the pair variables there are
reconstructed from a tomographically complete set of measurements, whereas here
both quadratures come from a single ratio of finite-time signals, with no
knowledge of the spectrum and no projection of the state. What we do
claim is the combination: that these standard objects, read through the ratio of
Eq.~(3) of the main text, give a coherence readout that is exact at finite time
for arbitrary Hamiltonians and unbiased under common-mode dephasing, and that
the old bound thereby acquires operational content --- it is the precision floor
of a specific, calibration-free readout protocol, and it disappears at
population balance. A two-parameter Cram\'er--Rao treatment going beyond the
Robertson form is left to future work (S8).

\begin{figure}[t]
\includegraphics[width=\textwidth]{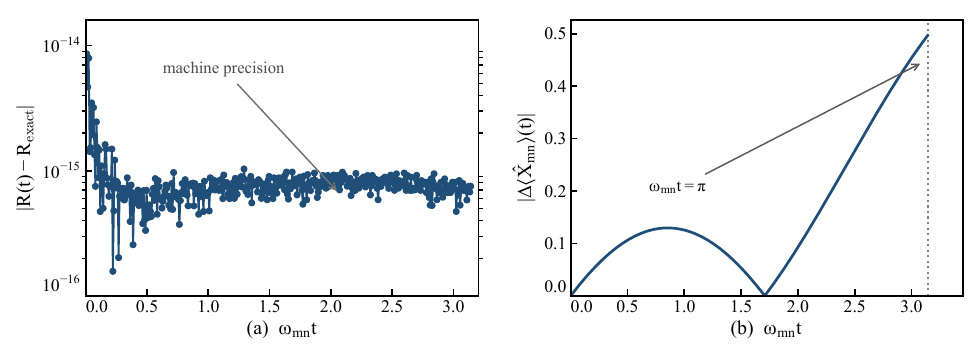}
\caption{Universality. An arbitrary three-level Hamiltonian with non-equidistant,
non-resonant levels drawn at random. (a) The deviation $|R(t)-R_{\rm exact}|$ of
the measured ratio from its exact value, for the real and imaginary parts
($R_{\rm exact}=\rho_{nm}/\rho_{r,nm}$), stays at machine precision for all $t$.
(b) The Hermitian readout $|\Davg{\hat X_{mn}}(t)|$ reaches order unity at
$t=\pi/\omega_{mn}$.}
\label{fig:general}
\end{figure}

For the trade-off we take a pure state on the pair,
$|\psi\rangle=c_m|E_m\rangle+c_n|E_n\rangle$ with complex amplitudes
$c_m,c_n$ obeying $|c_m|^2+|c_n|^2=1$. The information carried by the readout is
$S=2|c_mc_n|$, the same quantity that the main text writes as
$S_{mn}=2|\rho_{nm}|$, and $D$ is the disturbance it causes, i.e.\ one minus the
fidelity between the state before and after the readout.

\begin{figure}[t]
\includegraphics[width=\textwidth]{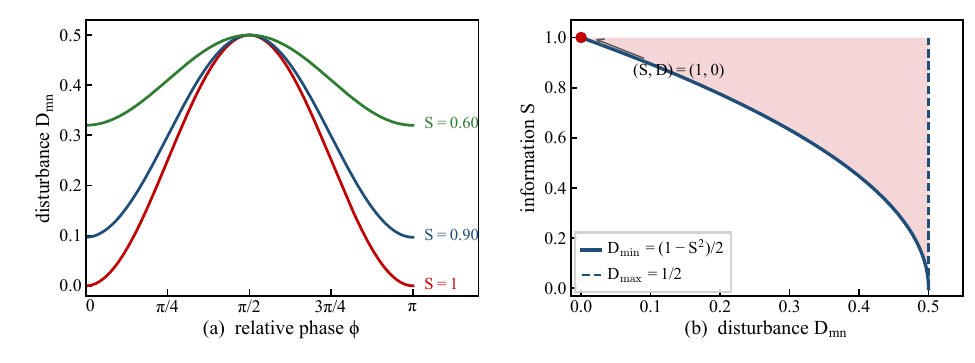}
\caption{Information--disturbance trade-off for a pure state on the pair
$(m,n)$. (a) Disturbance versus relative phase for three values of the
information $S=2|c_mc_n|$, labelled on the curves ($p_m=0.50,0.72,0.90$ for
$S=1,0.90,0.60$). (b) Attainable region: at given $S$ the disturbance lies
between $D_{\min}=(1-S^2)/2$ (solid) and $D_{\max}=1/2$ (dashed).}
\label{fig:tradeoff}
\end{figure}

Figure~\ref{fig:general} verifies Theorem~1 on an arbitrary Hamiltonian, with no
integrability, no equidistant levels and no conserved quantity beyond $\hat H$ itself.
Figure~\ref{fig:tradeoff} shows the trade-off region derived in the main text.

For the indirect, probe-mediated readout the disturbance acquires a time
dependence through the target--probe entanglement, and a closed form is
available. For a two-level atom $|\phi_a\rangle=a|g\rangle+b|e\rangle$, with
complex amplitudes $a,b$ obeying $|a|^2+|b|^2=1$, in a
vacuum cavity, the reduced-state infidelity $D_{\rm red}=1-F_{\rm red}$ obeys
$F_{\rm red}(t)=|a|^4+|a|^2|b|^2\sin^2(gt)+|b|^4\cos^2(gt)+2|a|^2|b|^2\cos(gt)$,
so $D_{\rm red}\simeq|b|^4(gt)^2$ and $S/D_{\rm red}\simeq1/|b|^2$ at short
times. Conditioning on the dominant branch (zero photons detected),
$1-F_{\rm cond}\simeq\tfrac12|a|^2|b|^4(gt)^4$ is of order $t^4$: the dominant
branch leaves the atom almost untouched while the information is carried by the
rare branch, which is the weak-measurement structure. For $|a|=|b|=1/\sqrt2$ the
time parameter is eliminated and
\begin{equation}
S=2D_{\rm red}(1-D_{\rm red}),
\end{equation}
with $S_{\max}=1/2$ at $gt=\pi/2$, where the atom is deterministically mapped to
$|g\rangle$. In both the direct and the indirect case the readout is a
deterministic, reversible map rather than a random projection.

Models with a conserved excitation number and a ladder structure --- the
Jaynes--Cummings, Dicke and Tavis--Cummings models --- are special cases in
which a natural observable already plays the role of the ruler. With the
excitation-number operator
$\hat N=\hat a^\dagger\hat a+\sum_j\hat\sigma_+^{(j)}\hat\sigma_-^{(j)}$
conserved, for a Fock field $|n\rangle$ of $n$ photons the photon-number change
factorizes exactly,
$\Davg{\hat a^\dagger\hat a}(t)=p\,P_e\,\sin^2(\Omega_{n,p}t)$, where
$\Omega_{n,p}=g\sqrt{(n+p)!/n!}$ is the Rabi frequency of the transition that
exchanges $p$ photons with the atomic ladder, $g$ the atom--photon coupling and
$P_e=|\langle e|\phi_a\rangle|^2$ the excited-state population of the atom,
giving a time-independent ratio with $P_e$ in place of $\rho_{nm}$. For the Rabi model the
counter-rotating terms couple each sector to more than two states, so the ratio
is not exactly $t$-independent at finite $t$; one then reverts to the
transition-operator readout of Theorem~1.

Algebraically, the transition operators form the standard basis of
$\mathfrak{gl}(d)$: $\hat A_{mn}\hat A_{kl}=\delta_{nk}\hat A_{ml}$ and
$[\hat A_{mn},\hat A_{kl}]=\delta_{nk}\hat A_{ml}-\delta_{lm}\hat A_{kn}$. The Hamiltonian
$\hat H=\sum_n E_n\hat A_{nn}$ lies in the Cartan subalgebra and the $\hat A_{mn}$ are its root
vectors, $[\hat H,\hat A_{mn}]=\omega_{mn}\hat A_{mn}$ with roots $\omega_{mn}$. The data ruler
is a root-space readout: it measures the coordinate $\rho_{nm}$ of the state
along the root direction $\hat A_{mn}$.

\section{S7. Classical limit}

Hats mark operators, so the classical quantities of this section carry none. In
Hamiltonian mechanics a phase-space function $A$ evolves as $dA/dt=\{A,H\}$
with the Poisson bracket $\{A,H\}=\sum_i(\partial_{\theta_i}A\,\partial_{I_i}H
-\partial_{I_i}A\,\partial_{\theta_i}H)$. For an integrable system the
action--angle variables $(I,\theta)$ give $H=H(I)$ and
$\theta(t)=\theta(0)+\omega(I)t$ with the frequency vector
$\omega(I)=\partial H/\partial I$, so $A(I,\theta)=\sum_kA_k(I)e^{ik\cdot\theta}$
for integer wave vectors $k$, and the Fourier modes are the classical
eigen-operators,
\begin{equation}
\{e^{ik\cdot\theta},H\}=i(k\cdot\omega)e^{ik\cdot\theta},
\end{equation}
in exact analogy with $[\hat H,\hat A_{mn}]=\omega_{mn}\hat A_{mn}$: the two
evolutions are $\dot{\hat A}=i[\hat H,\hat A]$ and $\dot A=\{A,H\}$, so the
correspondence is $i[\hat H,\cdot]\leftrightarrow\{\cdot,H\}$ and the
eigenvalues match as $\omega_{mn}\leftrightarrow k\cdot\omega$. (With the
opposite bracket order one gets
$\{H,e^{ik\cdot\theta}\}=-i(k\cdot\omega)e^{ik\cdot\theta}$.) Consequently
$\langle e^{ik\cdot\theta}\rangle(t)=e^{ik\cdot\omega t}\langle
e^{ik\cdot\theta}\rangle(0)$, where $\langle\cdot\rangle$ is the ensemble
average, and the ratio of the changes for two initial ensembles is exactly
time-independent, with the $k$-th Fourier component
$\rho_k=\langle e^{ik\cdot\theta}\rangle(0)$ read off against a reference. For
the harmonic oscillator of mass $m$, momentum $p$ and position $x$ the mode
$k=1$ is the complex amplitude $\alpha=x+ip/(m\omega)$, so the classical ruler
reads the initial phasor.

Three differences separate the quantum ruler from this limit. (i) Classical
mechanics has no measurement disturbance, so the
information--disturbance trade-off above is purely quantum. (ii) The precision
bound $\Delta\hat X_{mn}\Delta\hat Y_{mn}\geq|p_m-p_n|$ has no classical
counterpart, because $\{\cos k\theta,\sin k\theta\}=0$: the real and imaginary
parts commute classically and can be read simultaneously with arbitrary
precision. (iii) Universality over arbitrary $\hat H$ relies on the spectral
decomposition that every Hamiltonian possesses, whereas action--angle variables
exist only for integrable systems, so the classical ruler is restricted to
integrable dynamics.

\section{S8. Feasibility hierarchy and further directions}

The single practical constraint on all of the above is whether the readout of
$\hat A_{mn}=|E_m\rangle\langle E_n|$ can be implemented. It is useful to rank the
situations:

\begin{itemize}
\item[\textbf{L0}] \textbf{Natural observable, immediate.} When a conservation law
binds a readily measured observable to a transition operator, the ruler requires
no new hardware. This is the case for the Jaynes--Cummings, Dicke and
Tavis--Cummings models, where the photon number $\hat a^\dagger\hat a$ plays the
role of the ruler, and hence for Tavis--Cummings-type quantum batteries. All
ergotropy experiments of Sec.~S4 can be run at this level.
\item[\textbf{L1}] \textbf{Engineered readout, small systems.} For $d\lesssim8$ the
matrix elements of $\hat A_{mn}$ can be measured through an ancilla interferometric
readout or reconstructed from a small tomographic post-processing, on
superconducting circuits, trapped ions or NMR.
\item[\textbf{L2}] \textbf{Generally infeasible.} For a generic many-body system
the dimension grows exponentially and the individual transition operators are
not addressable. Here the ruler has only conceptual value, unless one works in
the local-probe setting \cite{Chen2025Probe} where the effective dimension stays
small.
\end{itemize}

Two directions follow naturally. First, the pair $(|\rho_{nm}|,\arg\rho_{nm})$
is a two-parameter estimation problem on the virtual qubit, and
Eq.~(5) of the main text is its Robertson-type bound; upgrading it to a
quantum Cram\'er--Rao or Holevo bound would connect the ruler to the active
question of measurement incompatibility in multiparameter estimation
\cite{Wang2026Multiparam}, with the notable feature that the bound becomes
trivial, and simultaneous precise readout becomes possible, at population
balance $p_m=p_n$. Second, the freedom from calibration suggests using the ruler
as the readout layer of probe-based Hamiltonian learning, where the inability to
calibrate the measurement basis is currently a structural obstacle
\cite{Chen2025Probe}.

\section{S9. Experimental proposal: coherent ergotropy of a Tavis--Cummings quantum battery}

This section turns the application of Sec.~S4 into a concrete proposal, so that
the claim ``conventional schemes systematically underestimate the coherent
contribution'' becomes a statement with a platform, a shot budget and error bars.
All numbers below come from a Monte Carlo simulation of the full protocol.

\paragraph{Platform.} Three transmons coupled to one cavity mode, i.e.\ the
Tavis--Cummings model in the symmetric Dicke subspace, operated on the
two-excitation manifold, which is three dimensional. With coupling
$g/2\pi=8$\,MHz and detuning $\Delta/2\pi=3$\,MHz the transition frequencies are
$f_{01}=24.582$, $f_{12}=26.358$ and $f_{02}=50.941$\,MHz; the smallest splitting
is 1.776\,MHz, so a 2\,$\mu$s photon-number record (200 points at 10\,ns) gives
0.5\,MHz resolution and resolves all three lines. The battery Hamiltonian is the
atomic energy, $\omega_a/2\pi=6$\,GHz, and the charged state is prepared by
letting a two-photon Fock state discharge into the atoms, with the charging time
chosen to maximise the stored energy ($\tau_c=19.5$\,ns). The charged state is
taken to be partially mixed, as a finite-bandwidth charging pulse would produce;
the mixing is fixed so that the coherent share of the ergotropy is 50.4\%, the
order reported for coherent quantum batteries \cite{Zhang2026QB}, and is not
tuned in favour of the conclusion. This is the L0 level of the feasibility
hierarchy of Sec.~S8: the cavity photon number is a natural observable, no
ancilla and no engineered transition operator are needed.

\paragraph{Protocol.} (i) Prepare the reference state $|n{=}2\rangle$ (two cavity
photons, atoms in the ground state), which is a coherent superposition in the
energy basis and hence a usable ruler. (ii) Prepare the charged state.
(iii) Let both evolve under the same $\hat H$ and record the cavity photon number.
(iv) Detect the three lines on the reference record by FFT. (v) Fit both
records on the same set of frequencies by linear least squares and take
the amplitude ratio. Because the fit is linear and identical for the two
records, the photon-number matrix element $\hat O_{mn}$, the dephasing factor, the
readout contrast and even the spectral-leakage factor cancel in the ratio.
Populations, which do not evolve and are therefore invisible to the ruler, are
supplied by a separate energy-basis measurement taking 10\% of the budget.

\paragraph{Results.} The error is quoted in units of the coherent contribution
$\WW(\hat\rho_0)-\WW(\DD(\hat\rho_0))$; 0 means exact recovery and $-1$ means the coherence is
entirely lost. Preparation coherence survival is $q=0.70$, $T_2=60\,\mu$s,
single-shot readout fidelity $F_{\rm ro}=0.97$, and the Hamiltonian used by the experimenter is
mis-calibrated by 2\%.

\begin{table}[h]
\caption{Recovery of the coherent ergotropy versus the total shot budget
(mean $\pm$ s.d., 120 repetitions).}
\begin{ruledtabular}
\begin{tabular}{cccc}
Total shots & Data ruler & Direct tomography & Single-state inversion\\
\hline
$3\times10^3$ & $1.93\pm0.93$ & $0.04\pm0.68$ & $0.99\pm0.94$\\
$1\times10^4$ & $0.95\pm0.82$ & $-0.20\pm0.35$ & $0.39\pm0.58$\\
$3\times10^4$ & $0.49\pm0.48$ & $-0.30\pm0.25$ & $0.13\pm0.39$\\
$1\times10^5$ & $0.17\pm0.24$ & $-0.33\pm0.12$ & $0.01\pm0.22$\\
$3\times10^5$ & $0.04\pm0.10$ & $-0.34\pm0.07$ & $-0.03\pm0.15$\\
$1\times10^6$ & $-0.005\pm0.051$ & $-0.343\pm0.041$ & $-0.041\pm0.073$\\
\end{tabular}
\end{ruledtabular}
\end{table}

The three curves differ in kind, not in degree. The ruler is unbiased and its
error falls as $1/\sqrt{N}$. Direct tomography saturates at $-34\%$: enlarging
the shot budget shrinks the error bar but not the offset, because the
preparation dephasing has removed part of the coherence and the estimate is
biased one-sidedly, always underestimating the quantum advantage.
Single-state inversion leaves $-4\%$, which is the 2\% calibration error
propagated through the fit.

Two robustness checks are worth recording. First, the line positions found
automatically on the reference record were $24.511$, $26.468$ and $48.987$\,MHz
against true values $24.582$, $26.358$ and $50.941$\,MHz, an error of nearly
2\,MHz on the third line; fitting both records on the detected rather than the
true frequencies changes the result from $0.028\pm0.095$ to $0.014\pm0.097$,
i.e.\ not at all. The ratio is immune to line-position error, again because one
and the same frequency basis is used for both states. Second, scanning the
calibration error from 1\% to 10\% leaves the ruler unchanged (between $-0.015$
and $+0.012$) while the inversion error grows monotonically to $-0.123$.

\paragraph{Cost.} Five per cent accuracy on the coherent contribution needs about
$3\times10^5$ shots, i.e.\ roughly 30\,s at a 10\,kHz repetition rate for one
charging time, and about ten minutes for a scan over twenty charging times.

\begin{table}[h]
\caption{Error budget. ``Common'' means identical for the unknown and the
reference state, which is the condition under which the ratio cancels it.}
\begin{ruledtabular}
\begin{tabular}{lcc}
Source & Common-mode? & Effect on the ruler\\
\hline
Preparation dephasing & yes & none (cancels)\\
Evolution dephasing ($T_2$) & yes & none (cancels)\\
Readout contrast $F_{\rm ro}$ & yes & none (constant term cancels)\\
Line-position error & same fit basis & none (cancels)\\
Hamiltonian mis-calibration & --- & none for the ratio spectrum\\
Cavity photon loss $\kappa$ & \textbf{no} & residual bias (main risk)\\
Populations & --- & invisible to the ruler, measured separately\\
\end{tabular}
\end{ruledtabular}
\end{table}

\paragraph{Limitations, stated honestly.} Cavity photon loss is included in
the full Lindblad simulation reported in Table~\ref{tab:kappa}: the states
evolve under $\LL=-i[\hat H,\cdot]+\kappa\,\Diss{\hat a}+\gamma_\phi\sum_j\Diss{\hat P_j}$,
where $\Diss{\hat L}\hat\rho=\hat L\hat\rho\hat L^\dagger
-\tfrac12\{\hat L^\dagger\hat L,\hat\rho\}$ is the Lindblad dissipator of the
jump operator $\hat L$, $\kappa$ the cavity decay rate, $\gamma_\phi$ the
transmon dephasing rate and $\hat P_j=|E_j\rangle\langle E_j|$ the projector on
level $j$; leakage into the one- and zero-excitation sectors is retained. Photon loss is the
one channel of this platform that violates the immunity criterion of
Theorem~3, because $\hat a^\dagger\hat a=\hat n$ does not commute with $\hat H_{\rm TC}$;
Sec.~S11 bounds the resulting bias by
$\kappa\|\hat n_{\rm off}\|/2\pi\Delta f_{\min}$ and evaluates it. The
conclusion is that the residual systematic is at most about $1\%$ of the
coherent contribution for $\kappa^{-1}\geq3\,\mu$s, and about $0.1\%$ for a
long-lived 3D cavity, in both cases below the statistical floor of Table~%
\ref{tab:kappa}. What loss does cost is signal amplitude, not accuracy: the
exponential decay of the signal is a common factor for the pair $(m,n)$ and
cancels in the ratio, so the standard deviations at $\kappa^{-1}=3$ and
$30\,\mu$s agree to within $10\%$.

We also stress that the ruler reads coherences; populations do not
evolve and must come from a separate energy-basis measurement, so the complete
ergotropy needs two batches of experiments, not one.

\section{S10. Two-parameter readout bound}
\label{sec:s10}

This section supplies the material behind Theorem~2 of the main text: the
quantum Fisher and incompatibility matrices of the level-pair qubit, the
Holevo bound and its comparison with the SLD bound, the attainable precision of
single-copy measurements, and the Monte-Carlo comparison of the ruler with
single-state inversion under a level-calibration error. All numbers were
obtained by numerical evaluation of the expressions derived below.

\subsection{S10.1 Setup}

Restrict the state to the pair $(m,n)$ and normalise, $P=p_m+p_n>0$:
\begin{equation}
\hat\varrho=\frac{\hat P_{mn}\hat\rho_0\hat P_{mn}}{P}
=\frac12\big(\mathbb I+\mathbf r\cdot\boldsymbol\sigma\big),\qquad
\mathbf r=(x,y,r_z),
\end{equation}
where $\hat P_{mn}=|E_m\rangle\langle E_m|+|E_n\rangle\langle E_n|$ is the
projector onto the pair and $\boldsymbol\sigma=(\hat\sigma_x,\hat\sigma_y,\hat\sigma_z)$
the Pauli vector acting on it. With $x+iy=2\rho_{nm}/P$ and
$r_z=(p_m-p_n)/P$, the vector $\mathbf r=(x,y,r_z)$ is the Bloch vector of the
pair and obeys $|\mathbf r|\leq1$ (equivalently $|\rho_{nm}|^2\leq p_mp_n$). We
write $s=1-|\mathbf r|^2$ for the purity
deficit and use the polar parameters $c=|\rho_{nm}|/P$ and
$\varphi=\arg\rho_{nm}$, the modulus and phase of the coherence,
so that $x=2c\cos\varphi$, $y=2c\sin\varphi$ and $s=1-4c^2-r_z^2$. The pair
populations $p_m,p_n$ (hence $P$ and $r_z$) are assumed known or measured
separately; the ruler estimates $x$ and $y$.

\subsection{S10.2 SLD operators and the quantum Fisher matrix}

The two parameters to be estimated are the Cartesian coordinates $x$ and $y$,
i.e.\ $i\in\{x,y\}$ throughout; the corresponding symmetric logarithmic
derivative (SLD) operators are $\hat L_x$ and $\hat L_y$. Being Hermitian on a
qubit they can be written as
$\hat L_i=l_0^{(i)}\mathbb I+\mathbf l^{(i)}\cdot\boldsymbol\sigma$ with real
scalar $l_0^{(i)}$ and real vector $\mathbf l^{(i)}$, and they are defined by
the SLD equation $\partial_i\hat\varrho=(\hat\varrho \hat L_i+\hat L_i\hat\varrho)/2$;
solving it gives
\begin{equation}
\mathbf l^{(i)}=\partial_i\mathbf r+
\frac{\mathbf r\cdot\partial_i\mathbf r}{s}\,\mathbf r,
\qquad
l_0^{(i)}=-\,\mathbf r\cdot\mathbf l^{(i)},
\label{eq:sld}
\end{equation}
the second relation being the gauge $\Tr(\hat\varrho \hat L_i)=0$. With
$\Tr(\hat\varrho \hat L_i\hat L_j)=-l_0^{(i)}l_0^{(j)}+\mathbf l^{(i)}\!\cdot\!\mathbf l^{(j)}
+i\,\mathbf r\cdot(\mathbf l^{(i)}\times\mathbf l^{(j)})$,
\begin{equation}
F_{ij}=\Real\Tr(\hat\varrho \hat L_i\hat L_j)
=\mathbf l^{(i)}\!\cdot\!\mathbf l^{(j)}
-(\mathbf r\!\cdot\!\mathbf l^{(i)})(\mathbf r\!\cdot\!\mathbf l^{(j)}),
\end{equation}
where $\Real$ and $\Imag$ denote the real and imaginary parts and $F$ is the
SLD (quantum) Fisher matrix of the pair.
Equation~\eqref{eq:sld} was cross-checked against a brute-force vectorised solve
of the SLD equation on random admissible states (agreement $<10^{-10}$).

\paragraph{Cartesian parameters.} With $\mathbf e_x$ and $\mathbf e_y$ the
Cartesian unit vectors in the equatorial plane, $\partial_x\mathbf r=\mathbf e_x$
and $\partial_y\mathbf r=\mathbf e_y$, hence
\begin{equation}
F_{ij}=\delta_{ij}+\frac{r_ir_j}{s},\qquad i,j\in\{x,y\},
\qquad
\tr F^{-1}=2-\frac{x^2+y^2}{1-r_z^2}.
\label{eq:Fcart}
\end{equation}

\paragraph{Polar parameters.} Let $\mathbf e_c=(\cos\varphi,\sin\varphi,0)$ and
$\mathbf e_\varphi=(-\sin\varphi,\cos\varphi,0)$ be the radial and tangential
unit vectors of the in-plane polar coordinate system, with
$\mathbf e_c\perp\mathbf e_\varphi$ and both orthogonal to $\mathbf e_z$. Then
$\partial_c\mathbf r=2\mathbf e_c$ (radial),
$\partial_\varphi\mathbf r=2c\,\mathbf e_\varphi$ (tangential), so
$\mathbf r\cdot\partial_\varphi\mathbf r=0$ and $F_{c\varphi}=0$: the Fisher
matrix is diagonal,
\begin{equation}
F_{cc}=4+\frac{16c^2}{s},\qquad F_{\varphi\varphi}=4c^2,\qquad F_{c\varphi}=0 .
\label{eq:Fpolar}
\end{equation}
Modulus and phase are therefore statistically uncorrelated at the SLD level.
Two consistency checks: (i) at maximal coherence $p_m=p_n$ ($r_z=0$) and
$2c=1$, Eq.~\eqref{eq:Fpolar} gives $F_{\varphi\varphi}=1$, the standard
single-qubit phase-estimation QFI; (ii) $F_{cc}$ diverges as $s\to0$ because on
the pure-state manifold $c$ and $r_z$ are not independent ($4c^2+r_z^2=1$), so
the radial direction is not an admissible variation --- the divergence is a
statement about the parameterisation, not about the readout.

\subsection{S10.3 Incompatibility matrix}

The SLDs are Hermitian, so $[\hat L_i,\hat L_j]$ is anti-Hermitian and its expectation is
purely imaginary:
\begin{equation}
D_{ij}\equiv\Tr\big(\hat\varrho[\hat L_i,\hat L_j]\big)
=2i\,\mathbf r\cdot\big(\mathbf l^{(i)}\times\mathbf l^{(j)}\big).
\end{equation}
Since $\mathbf r\cdot(\mathbf e_x\times\mathbf r)=\mathbf r\cdot(\mathbf r\times
\mathbf e_y)=0$, only the $\mathbf e_x\times\mathbf e_y$ term survives, giving
\begin{equation}
|D_{xy}|=2|r_z|,\qquad |D_{c\varphi}|=8c|r_z| .
\label{eq:Dmat}
\end{equation}
Both are proportional to the population imbalance
$|p_m-p_n|/P$, the same quantity that sets the Robertson floor
$\Delta \hat X_{mn}\Delta \hat Y_{mn}\geq|p_m-p_n|$ of the main text: the two statements
have a common origin in $[\hat X_{mn},\hat Y_{mn}]=2i\hat Z_{mn}$. Numerical agreement with
Eq.~\eqref{eq:Dmat} is to machine precision.

\subsection{S10.4 Holevo bound and proof of Theorem~2}

$D_{ij}=0$ is the weak commutativity condition, necessary for the SLD
Cram\'er--Rao bound to be attainable but not sufficient. The attainable limit is
set by the Holevo bound \cite{Holevo2011,Suzuki2016}. For a two-parameter qubit
model with unit weight matrix it reads
\begin{equation}
C^{\rm H}=\min_{\hat{\mathbf M}}\Big\{\tr\Real Z[\hat{\mathbf M}]+\tr\big|\Imag Z[\hat{\mathbf M}]\big|
\Big\},\quad Z_{ij}[\hat{\mathbf M}]=\Tr(\hat\varrho \hat M_i\hat M_j),
\end{equation}
minimised over pairs of Hermitian operators $\hat M_1,\hat M_2$ on the pair
qubit --- not to be confused with the quadratures $\hat X_{mn},\hat Y_{mn}$ ---
subject to $\Tr(\hat\varrho \hat M_i)=0$ and
$\Tr(\partial_j\hat\varrho\,\hat M_i)=\delta_{ij}$. Writing
$\hat M_i=\alpha_i\mathbb I+\boldsymbol\beta_i\cdot\boldsymbol\sigma$ with real
$\alpha_i$ and real vector $\boldsymbol\beta_i$, the
constraints force $\boldsymbol\beta_1=(1,0,b_1)$,
$\boldsymbol\beta_2=(0,1,b_2)$, $\alpha_i=-\mathbf r\cdot\boldsymbol\beta_i$
with $b_1,b_2\in\mathbb R$ the only free parameters, and
\begin{equation}
C^{\rm H}=\min_{b_1,b_2}\Big[\,2+b_1^2+b_2^2-(x+r_zb_1)^2-(y+r_zb_2)^2
+2\big|r_z-xb_1-yb_2\big|\,\Big].
\label{eq:CH}
\end{equation}
(As a check, the one-parameter version returns $C^{\rm H}=1/F_{xx}$ to
$10^{-16}$--$10^{-8}$.)

\paragraph{Theorem 2.} Let $\mathcal{G}\equiv C^{\rm H}-\tr F^{-1}$ with
$\tr F^{-1}$ from Eq.~\eqref{eq:Fcart}, $U=x^2+y^2>0$ and $s=1-r_z^2-U>0$. Then
$\mathcal{G}=0$ if $r_z=0$, and $\mathcal{G}\geq r_z^2s^2/(1-r_z^2)>0$ if
$r_z\neq0$.

\paragraph{Proof.} At $r_z=0$, Eq.~\eqref{eq:CH} becomes
$2+b_1^2+b_2^2-x^2-y^2+2|xb_1+yb_2|\geq 2-x^2-y^2$, with equality at
$b_1=b_2=0$; and $\tr F^{-1}=2-(x^2+y^2)$ by Eq.~\eqref{eq:Fcart}. For
$r_z\neq0$ put $w=xb_1+yb_2$, so that $b_1^2+b_2^2\geq w^2/U$ by
Cauchy--Schwarz --- saturated by $(b_1,b_2)\parallel(x,y)$, so the bound is
attained --- and
\begin{equation}
\mathcal{G}\geq \frac{w^2(1-r_z^2)}{U}-2r_zw+2|r_z-w|+\frac{Ur_z^2}{1-r_z^2},
\label{eq:Gw}
\end{equation}
with equality. Writing $\zeta=|r_z|$ for the (scalar) imbalance, so that
$1-r_z^2=1-\zeta^2$ (the gap is even in $r_z$, since
$(b_1,b_2)\to(-b_1,-b_2)$ compensates $r_z\to-r_z$ in Eq.~\eqref{eq:CH}), the
right-hand side of Eq.~\eqref{eq:Gw} is convex and piecewise quadratic in~$w$.
On $w>\zeta$ its derivative $2(1-\zeta^2)w/U-2\zeta+2$ is positive, so that
branch is minimised at the kink $w=\zeta$, giving
$\mathcal{G}=\zeta^2s^2/[U(1-\zeta^2)]$. On $w\leq\zeta$ the stationary point
$w_\ast=U(1+\zeta)/(1-\zeta^2)$ lies in the branch only when
$U\leq\zeta(1-\zeta)$; then it, and not the kink, is the minimum, giving
$\mathcal{G}=2\zeta-U(1+2\zeta)/(1-\zeta^2)$. Hence, using
$U\leq\zeta(1-\zeta)\Leftrightarrow s\geq1-\zeta$,
\begin{equation}
\mathcal{G}=\begin{cases}
2\zeta-\dfrac{U(1+2\zeta)}{1-\zeta^2}, & s\geq 1-\zeta,\\[2ex]
\dfrac{\zeta^2s^2}{U(1-\zeta^2)}, & s\leq 1-\zeta .
\end{cases}
\label{eq:Gexact}
\end{equation}
Both branches are strictly positive: the first decreases in $U$ and equals
$\zeta/(1+\zeta)$ at the branch point; the second is manifestly positive. The
branch-free bound quoted in the main text,
$\mathcal{G}\geq\zeta^2s^2/(1-\zeta^2)=r_z^2s^2/(1-r_z^2)$, follows because
$U<1$ in the second branch, and in the first because
$2\zeta(1-\zeta^2)-U(1+2\zeta)-\zeta^2(1-\zeta^2-U)^2$ is concave in $U$ and
non-negative at both endpoints $U=0$ (where it is
$2\zeta(1-\zeta^2)-\zeta^2(1-\zeta^2)^2\geq0$) and $U=\zeta(1-\zeta)$ (where it
is $\zeta(1-\zeta)[1-\zeta(1-\zeta)]\geq0$).

\paragraph{Remark.} An earlier version of this proof asserted that both branches of
Eq.~\eqref{eq:Gw} are minimised at $w=\zeta$ and hence claimed the stronger
bound $\mathcal{G}\geq r_z^2s^2/[U(1-r_z^2)]$. That is false in the first
branch: at $U<\zeta(1-\zeta)$ the interior stationary point wins, and the claimed
bound can exceed the true gap by more than an order of magnitude (e.g.\
$\zeta=0.5$, $U=0.01$: true $\mathcal{G}=0.9733$ against a claimed $18.25$).
The qualitative conclusion --- $\mathcal{G}=0$ exactly at $r_z=0$ and
$\mathcal{G}>0$ otherwise --- is unaffected.

Table~\ref{tab:holevo} shows the numbers. The gap vanishes identically at
$r_z=0$ and grows with the imbalance.

\begin{table}[h]
\caption{Holevo bound vs.\ the SLD bound $\tr F^{-1}$ for the joint estimation
of $(x,y)$, per copy, at fixed in-plane radius $|\mathbf r_\parallel|$ and
$\varphi=0$. Every entry is the exact value of Eq.~\eqref{eq:Gexact}, obtained
by minimising Eq.~\eqref{eq:CH} in closed form. The last two rows sit in the
interior branch, where the earlier $\propto1/(x^2+y^2)$ claim would have
returned $\mathcal{G}=4.20$ and $18.25$ instead of $0.89$ and $0.97$.}
\label{tab:holevo}
\begin{ruledtabular}
\begin{tabular}{cccccc}
$|\mathbf r_\parallel|$ & $r_z$ & $\tr F^{-1}$ & $C^{\rm H}$ & $\mathcal{G}$
 & branch\\
\hline
0.60 & 0.00 & 1.6400 & 1.6400 & 0.00000 & ---\\
0.60 & 0.10 & 1.6364 & 1.6475 & 0.01114 & kink\\
0.60 & 0.20 & 1.6250 & 1.6667 & 0.04167 & kink\\
0.60 & 0.35 & 1.5897 & 1.6936 & 0.10385 & kink\\
0.60 & 0.50 & 1.5200 & 1.6608 & 0.14083 & kink\\
0.80 & 0.00 & 1.3600 & 1.3600 & 0.00000 & ---\\
0.80 & 0.10 & 1.3535 & 1.3555 & 0.00193 & kink\\
0.80 & 0.20 & 1.3333 & 1.3400 & 0.00667 & kink\\
0.80 & 0.35 & 1.2707 & 1.2830 & 0.01230 & kink\\
0.80 & 0.50 & 1.1467 & 1.1530 & 0.00630 & kink\\
\hline
0.20 & 0.50 & 1.9467 & 2.8400 & 0.89333 & interior\\
0.10 & 0.50 & 1.9867 & 2.9600 & 0.97333 & interior\\
\end{tabular}
\end{ruledtabular}
\end{table}

\subsection{S10.5 Single-copy measurements: a second, state-independent cost}

Theorem~2 concerns the asymptotic limit with possibly collective measurements.
An experiment that measures one copy at a time pays an additional cost. We
minimised $\tr\mathcal{I}^{-1}$ over general $K$-outcome qubit POVMs
$\{\hat\Pi_k\}$, where $\mathcal{I}$ is the classical Fisher information matrix
of the POVM, $K$ the number of outcomes and
$\hat\Pi_k=a_k(\mathbb I+\mathbf m_k\cdot\boldsymbol\sigma)/2$ with $a_k\geq0$,
$\sum_ka_k=2$, $|\mathbf m_k|\leq1$, $\sum_ka_k\mathbf m_k=0$, using
multi-start L-BFGS-B and rejecting any candidate that violates the constraints
to $10^{-5}$. The reported values are therefore upper bounds on the attainable
covariance, i.e.\ conservative.

\begin{table}[h]
\caption{Single-copy attainability. $\tr\mathcal{I}^{-1}$ of the best POVM found,
compared with the SLD bound and with the naive strategy of splitting the
samples equally between the two quadratures.}
\label{tab:singlecopy}
\begin{ruledtabular}
\begin{tabular}{ccccc}
$|\mathbf r_\parallel|$ & $r_z$ & $\tr F^{-1}$ & best POVM & ratio\\
\hline
0.60 & 0.00 & 1.6400 & 3.2408 & 1.98\\
0.60 & 0.35 & 1.5897 & 3.1261 & 1.97\\
0.80 & 0.00 & 1.3600 & 2.5600 & 1.88\\
0.80 & 0.35 & 1.2707 & 2.3113 & 1.82\\
0.80 & 0.50 & 1.1467 & 1.9127 & 1.67\\
\end{tabular}
\end{ruledtabular}
\end{table}

Two conclusions follow, and both matter for reading the main text correctly.
First, the single-copy cost is a factor $1.7$--$2.0$ over the SLD bound and is
not removed by population balance: at $r_z=0$ it is still $1.88$--$1.98$.
It comes from $[\hat X_{mn},\hat Y_{mn}]=2i\hat Z_{mn}$, i.e.\ from the impossibility of
reading both quadratures of one copy, and is state-independent --- unlike the
cost of Theorem~2, which is switched by $r_z$. Splitting the samples equally
between the two quadratures gives exactly $2.00$ at $r_z=0$, so the elaborate
POVM buys at most a few per cent: there is no practical reason not to use the
simple split. Second, neither the Robertson floor nor Theorem~2 sees this
factor; a joint readout is never free, only asymptotically so.

\subsection{S10.6 Calibration-free readout: ruler against single-state inversion}

Theorem~1 makes the ratio independent of $\omega_{mn}$, so the ruler needs no
level calibration; inverting a single-state signal does. To quantify this we
simulated the pair subspace with $\hat H=\frac\omega2\hat\sigma_z$,
$\omega/2\pi=5$\,MHz, an unknown state with $|\rho_{nm}|/P=0.36$,
$\arg\rho_{nm}=0.7$, $r_z=0.10$, and a reference state with in-plane radius
$0.70$ and phase $0$. The experimenter believes
$\omega_{\rm est}=\omega(1+\epsilon)$ and takes $t_\star=\pi/\omega_{\rm est}$; both
schemes measure $\hat X_{mn}$ and $\hat Y_{mn}$ at $t=0$ and $t=t_\star$ and use the same
total shot budget $N$ on the unknown state. The ruler additionally spends
$N_{\rm ref}$ shots once to calibrate the reference signal.

\begin{table}[h]
\caption{MSE of $\rho_{nm}$ at $N=4\times10^4$, $N_{\rm ref}=2\times10^4$,
400 repetitions, as a function of the level-calibration error.}
\label{tab:calib}
\begin{ruledtabular}
\begin{tabular}{cccc}
calibration error & data ruler & single-state inversion & ratio\\
\hline
0\%   & 2.86e-5 & 1.81e-5 & 0.63\\
1\%   & 2.77e-5 & 5.09e-5 & 1.84\\
3\%   & 2.65e-5 & 2.87e-4 & 10.8\\
10\%  & 2.93e-5 & 2.65e-3 & 90.4\\
\end{tabular}
\end{ruledtabular}
\end{table}

The first row is the honest cost of the ruler: at perfect calibration the
inversion is better (ratio $0.63$), because the ruler pays for the finite
calibration of its reference. Any calibration error inverts the comparison.

Table~\ref{tab:shots} shows that the two errors differ in kind, not in size.
The inversion saturates at a floor of $2.7\times10^{-4}$ that no shot budget
removes --- a bias, exactly as in Sec.~S4 --- whereas the ruler decreases
roughly as $1/N$ until it meets the floor set by its own one-off calibration.

\begin{table}[h]
\caption{Shot scaling at $3\%$ calibration error, and the ruler floor set by
$N_{\rm ref}$ (at $N=3.2\times10^5$).}
\label{tab:shots}
\begin{ruledtabular}
\begin{tabular}{ccc}
$N$ & data ruler & inversion\\
\hline
2.0e4  & 4.45e-5 & 2.99e-4\\
8.0e4  & 1.82e-5 & 2.78e-4\\
3.2e5  & 1.22e-5 & 2.74e-4\\
\hline
$N_{\rm ref}$ & data ruler & inversion\\
\hline
2.0e4  & 1.22e-5 & 2.74e-4\\
1.0e5  & 4.47e-6 & 2.75e-4\\
1.0e6  & 2.53e-6 & 2.72e-4\\
\end{tabular}
\end{ruledtabular}
\end{table}

The lower block is the point of the ruler as a ruler: its calibration is
a one-off cost, amortised over every subsequent unknown state, so raising
$N_{\rm ref}$ lowers the floor without raising the per-experiment cost.

\subsection{S10.7 What is not claimed}

The qubit quantum Fisher matrix, the SLD incompatibility matrix and the
construction of the Holevo bound are all standard
\cite{Holevo2011,Suzuki2016,Ragy2016,Szczykulska2016}; we claim no novelty for
them. The observation we do claim is the switch: for this readout the
entire gap between the Holevo and SLD bounds is controlled by
$r_z=(p_m-p_n)/(p_m+p_n)$, vanishing if and only if the populations are equal,
with the explicit lower bound of Theorem~2. The second, state-independent
single-copy factor of $\approx2$ is reported as a limitation of the readout, not
as an advantage.

\section{S11. Which noise channels the ruler is immune to}

Theorem~3 of the main text rests on an algebraic criterion. This section proves
it in full, tests it channel by channel, bounds the error when it fails, and
evaluates that bound on the platform of Sec.~S9.

\subsection{S11.1 Criterion and proof}

Let $\LL$ be the generator, so that $\hat\rho(t)=e^{\LL t}\hat\rho_0$ and
$\langle\hat A\rangle(t)=\Tr[\hat A e^{\LL t}\hat\rho_0]=\Tr[(e^{\LL^\dagger t}\hat A)\hat\rho_0]$. If
$\hat A_{mn}$ is an eigen-operator of the adjoint generator,
$\LL^\dagger(\hat A_{mn})=\lambda_{mn}\hat A_{mn}$, then
$\langle \hat A_{mn}\rangle(t)=e^{\lambda_{mn}t}\langle \hat A_{mn}\rangle(0)$ and
\begin{equation}
\Davg{\hat A_{mn}}(t,\hat\rho_0)=\rho_{nm}\big(e^{\lambda_{mn}t}-1\big),
\end{equation}
so the bracket --- which depends only on the generator, not on the state ---
cancels in the ratio. Conversely, if $\hat A_{mn}$ is not an eigen-operator, the
evolution of $\langle \hat A_{mn}\rangle$ draws on matrix elements other than
$\rho_{nm}$, and no ratio of two such signals can return $\rho_{nm}/\rho_{r,nm}$
for all $\hat\rho_0$. The eigen-operator property is therefore necessary as well as
sufficient.

For a Lindbladian with jump operators $\hat L_\alpha$ \cite{Breuer2002},
\begin{equation}
\LL^\dagger(\hat A)=i[\hat H,\hat A]+\sum_\alpha\Big(\hat L_\alpha^\dagger\hat A\hat L_\alpha
-\tfrac12\{\hat L_\alpha^\dagger\hat L_\alpha,\hat A\}\Big).
\end{equation}
Write $\hat L=\sum_{jk}c_{jk}|E_j\rangle\langle E_k|$ with matrix elements $c_{jk}=\langle E_j|\hat L|E_k\rangle$, and $\hat G\equiv\hat L^\dagger\hat L$ with matrix elements $G_{jk}=\langle E_j|\hat G|E_k\rangle$.
Then
\begin{align}
\tfrac12\{\hat G,\hat A_{mn}\}&=\tfrac12(G_{mm}+G_{nn})\hat A_{mn}\notag\\
&\quad+\tfrac12\sum_{a\neq m}G_{am}|E_a\rangle\langle E_n|
+\tfrac12\sum_{b\neq n}G_{nb}|E_m\rangle\langle E_b|,\\
\hat L^\dagger\hat A_{mn}\hat L&=\sum_{kk'}c^*_{mk}c_{nk'}|E_k\rangle\langle E_{k'}|.
\end{align}
The first expression is proportional to $\hat A_{mn}$ for every pair iff $G$ is
diagonal in the energy basis, i.e.\ $[\hat G,\hat H]=[\hat L^\dagger\hat L,\hat H]=0$. The second requires
more care, and it is here that a looser statement made in an earlier version has
to be corrected. It is not enough that $\hat L$ be monotone (purely lowering,
purely raising, or diagonal).

\paragraph{Counterexample.} Take a three-level ladder $E_0<E_1<E_2$ and the
single jump operator
\begin{equation}
\hat L=|E_0\rangle\langle E_1|+|E_1\rangle\langle E_2| .
\end{equation}
Then $\hat L$ is purely lowering, and
$\hat L^\dagger\hat L=|E_1\rangle\langle E_1|+|E_2\rangle\langle E_2|$ is diagonal, so
both halves of the looser criterion hold. Yet for the pair $(m,n)=(0,1)$,
\begin{equation}
\hat L^\dagger\hat A_{01}\hat L=|E_1\rangle\langle E_2|\neq0 ,
\end{equation}
because row $0$ of $\hat L$ is non-empty at column $1$ and row $1$ at column $2$,
so the coefficient $c^*_{mk}c_{nk'}$ survives with $(k,k')=(1,2)$. Hence
$\LL^\dagger(\hat A_{01})$ is not proportional to $\hat A_{01}$: numerically the
eigen-operator residual is $1.0$, and the ruler ratio for that pair is wrong by
$42\%$, $90\%$ and $53\%$ at $t=0.4,0.9,1.7$ (arbitrary units), whereas the
remaining pairs are exact. The same physics written as two jump
operators, $\hat L_1=|E_0\rangle\langle E_1|$ and
$\hat L_2=|E_1\rangle\langle E_2|$, gives a residual of exactly $0$ for every pair:
the cross terms are an artefact of putting two transitions inside one $\hat L$,
because $\sum_\alpha \hat L_\alpha^\dagger\hat A\hat L_\alpha$ never mixes different
$\alpha$.

The correct sufficient condition on each jump operator separately is therefore
\begin{equation}
\hat L_\alpha\ \text{diagonal in the energy basis,}\quad\text{or}\quad
\hat L_\alpha\propto|E_j\rangle\langle E_k|\ \text{with}\ j\neq k .
\label{eq:jumpterm}
\end{equation}
Indeed, if $\hat L$ is diagonal then $c_{mk}=\delta_{mk}c_{mm}$ and the whole sum
collapses to $c^*_{mm}c_{nn}\hat A_{mn}\propto \hat A_{mn}$; if $L\propto|E_j\rangle
\langle E_k|$ with $j\neq k$ then $c_{mk}$ and $c_{nk'}$ cannot both be nonzero
for $m\neq n$, so the sum vanishes. Together with $[\hat L^\dagger\hat L,\hat H]=0$ this
covers phase damping and spontaneous emission, and for a non-degenerate $\hat H$ no
other single jump operator does: if two rows of $\hat L$ carry off-diagonal weight
the sum produces $|E_k\rangle\langle E_{k'}|$ with $(k,k')\neq(m,n)$, and if one
row carries it at two columns then $\hat L^\dagger\hat L$ acquires off-diagonal
elements.
When the condition holds,
\begin{equation}
\lambda_{mn}=i\omega_{mn}+\sum_\alpha\Big[c^{(\alpha)*}_{mm}c^{(\alpha)}_{nn}
-\tfrac12\big(\Gamma^{(\alpha)}_m+\Gamma^{(\alpha)}_n\big)\Big],\qquad
\Gamma^{(\alpha)}_m=(\hat L_\alpha^\dagger \hat L_\alpha)_{mm},
\end{equation}
with $c^{(\alpha)}_{jk}=\langle E_j|\hat L_\alpha|E_k\rangle$. Two special cases are
worth recording, because the compact formula hides them. For emission,
$c_{mm}=c_{nn}=0$ and $\lambda_{mn}=i\omega_{mn}-\tfrac12(\Gamma_m+\Gamma_n)$
--- the coherence decays at the mean of the two level decay rates. For
dephasing with $\hat L_\alpha=\sum_jc^{(\alpha)}_j|E_j\rangle\langle E_j|$ the two
pieces combine into
$-\tfrac12\sum_\alpha|c^{(\alpha)}_m-c^{(\alpha)}_n|^2$; for the common choice
$\hat L=\sqrt{\gamma_\phi}\,\hat H$ this is the familiar
$-\tfrac{\gamma_\phi}{2}\omega_{mn}^2$. Writing only
$\lambda_{mn}=i\omega_{mn}-\tfrac12(\Gamma_m+\Gamma_n)$, as an earlier version
did, is wrong for that channel: it would give
$-\tfrac{\gamma_\phi}{2}(E_m^2+E_n^2)$ instead of
$-\tfrac{\gamma_\phi}{2}(E_m-E_n)^2$. The immunity itself is unaffected,
since $\lambda_{mn}$ cancels in the ratio whatever its value.

\subsection{S11.2 Channel-by-channel test}

We evaluate, for each channel, the worst-case relative failure of the
eigen-operator property over all pairs,
$\max_{m\neq n}\|\LL^\dagger(\hat A_{mn})-\lambda \hat A_{mn}\|/\|\LL^\dagger(\hat A_{mn})\|$,
minimised over $\lambda$.

\begin{table}[h]
\caption{Eigen-operator residual for representative channels. ``Qutrit'': a
three-level system with splittings $3.0$ and $7.7$~MHz. ``TC'': the
Tavis--Cummings model of Sec.~S9. Residuals at the $10^{-15}$ level are machine
precision, i.e.\ exact immunity.}
\begin{ruledtabular}
\begin{tabular}{lcc}
Channel & System & Residual \\
\hline
Energy-basis dephasing & qutrit & $0$ \\
Spontaneous emission (3 lowering channels) & qutrit & $0$ \\
Dephasing $+$ emission & qutrit & $0$ \\
Energy-basis dephasing & TC & $1.7\times10^{-15}$ \\
Cavity decay, $\kappa^{-1}=3\ \mu$s & TC & $1.07\times10^{-2}$ \\
Collective atomic decay $\hat J_-$ & TC & $6.4\times10^{-4}$ \\
\end{tabular}
\end{ruledtabular}
\end{table}

Two facts deserve emphasis. First, amplitude damping is immune: for
$\hat L=|E_j\rangle\langle E_k|$ one has $\hat L^\dagger\hat L=|E_k\rangle\langle E_k|$,
manifestly commuting with $\hat H$, and the coherence simply decays at the mean of
the two level decay rates, $\tfrac12(\Gamma_m+\Gamma_n)$. Immunity is therefore
not a statement about dephasing alone. Second, cavity decay fails because
the photon number is not a good quantum number: $\hat L=\sqrt\kappa\,\hat a$ gives
$\hat L^\dagger\hat L=\kappa\hat n$, whose off-diagonal elements in the polariton energy
basis reach $0.70$ in the TC model. The loss channel is a single coherent one
serving several transitions simultaneously, which is precisely the case the
criterion excludes.

\subsection{S11.3 Bound on the residual bias}

When the criterion fails only weakly, first-order perturbation theory gives
\begin{equation}
\varepsilon\ \equiv\ \Big|\frac{\rho_{nm}^{\rm est}/\rho_{r,nm}^{\rm est}}
{\rho_{nm}/\rho_{r,nm}}-1\Big|
\ \lesssim\ \sum_\alpha\frac{\|(\hat L_\alpha^\dagger \hat L_\alpha)_{\rm off}\|}
{2\pi\,\Delta f_{\min}},
\label{eq:epsbound}
\end{equation}
where $\Delta f_{\min}$ is the smallest spacing in transition frequency
between the sectors that the channel connects, and the numerator is the largest
off-diagonal element of $\hat L_\alpha^\dagger \hat L_\alpha$ in the energy basis. The
$2\pi$ converts transition frequencies (in Hz) to angular frequencies.

\begin{table}[h]
\caption{TC battery, noiseless. Bias of the extracted ratio and the resulting
systematic error on the ergotropy, in units of the coherent contribution
$\WW(\hat\rho_0)-\WW(\DD(\hat\rho_0))$. The $\kappa^{-1}=\infty$ row is the protocol floor
set by spectral leakage of a $2\ \mu$s record, not by loss.}
\begin{ruledtabular}
\begin{tabular}{cccc}
$\kappa^{-1}$ ($\mu$s) & bias, pair 01 & bias, pair 02 & ergotropy error \\
\hline
\label{tab:kappa}
$\infty$ & $2.1\times10^{-4}$ & $3.1\times10^{-3}$ & $-0.000$ \\
300 & $2.6\times10^{-4}$ & $3.6\times10^{-3}$ & $-0.000$ \\
100 & $3.6\times10^{-4}$ & $4.7\times10^{-3}$ & $-0.001$ \\
30 & $7.2\times10^{-4}$ & $8.7\times10^{-3}$ & $-0.001$ \\
10 & $1.8\times10^{-3}$ & $2.0\times10^{-2}$ & $-0.003$ \\
3 & $6.1\times10^{-3}$ & $6.5\times10^{-2}$ & $-0.006$ \\
1 & $2.3\times10^{-2}$ & $2.2\times10^{-1}$ & $\phantom{-}0.000$ \\
\end{tabular}
\end{ruledtabular}
\end{table}

With $\Delta f_{\min}=1.516$~MHz (the gap between $f_{12}=26.36$~MHz and the
$27.875$~MHz transition of the leakage sector $N=1$) and
$\|\hat n_{\rm off}\|=0.70$, Eq.~\eqref{eq:epsbound} predicts
$2.5\times10^{-3}$, $7.4\times10^{-3}$ and $2.5\times10^{-2}$ at
$\kappa^{-1}=30$, $10$ and $3\ \mu$s; subtracting the protocol floor, the
measured biases are larger by a factor $2.3$--$2.5$. The bound is therefore
correct in order of magnitude and conservative, and the bias is linear in
$\kappa$ over two decades.

Two honest qualifications. (i) The row $\kappa^{-1}=\infty$ is nonzero:
$3.1\times10^{-3}$ comes from spectral leakage of the finite record, a property
of the fitting protocol rather than of the noise. Windowing or a longer record
reduces it. (ii) The bias in the ratio does not propagate one-to-one into
the ergotropy: the largest bias sits on the pair $(0,2)$, whose $|\rho_{02}|$ is
an order of magnitude smaller than the other two coherences for this charged
state. That is a property of this state, not a theorem; for a state with a large
$\rho_{02}$ the ergotropy error should be expected at the $10^{-2}$ level. The
conservative statement is that for $\kappa^{-1}\geq3\ \mu$s the residual
systematic is at most about $1\%$ of the coherent contribution, against a
statistical uncertainty of $\pm2\%$ at $3\times10^{5}$ shots (Sec.~S9).

A related point deserves to be stated explicitly, because it is easy to get
wrong: photon loss within the $2\ \mu$s record removes about half the population
at $\kappa^{-1}=3\ \mu$s, yet it barely degrades the statistics. Comparing the
Monte Carlo columns of Sec.~S9 at $\kappa^{-1}=3$ and $30\ \mu$s, the standard
deviations agree to within $10\%$. The exponential loss of signal amplitude is a
common factor for the pair $(m,n)$ and cancels in the ratio, so leakage costs a
constant factor in amplitude, not a systematic error.

\subsection{S11.4 A platform with no non-common-mode channel}

The bound above is reassuring but it is still a bound: on circuit-QED hardware
one must always argue that a residual is small. There exists a family of
platforms where no such argument is needed, because the criterion holds exactly
and no channel violating it is present.

Take three ground hyperfine--Zeeman levels of a trapped ion (or, equivalently, a
spin-1 nucleus in NMR), with transition frequencies $3.0$, $4.7$ and $7.7$~MHz.
(Circuit-QED parameters quoted in Sec.~S9 follow
Refs.~\cite{Blais2021,Paik2011}.)
All three levels are ground states, so there is no spontaneous emission at all;
had there been, the decay channels would be lowering operators between energy
eigenstates and would satisfy the criterion anyway. The only noise is dephasing,
which is diagonal and therefore immune. There is no cavity. The observable is
$\hat O=|v\rangle\langle v|$ with $v=(1,1,1)/\sqrt3$, realised by an analysis pulse
$\hat U_A$ with $\hat U_A|v\rangle=|0\rangle$ followed by a population readout; this
equalises all three $|O_{mn}|$ at $1/3$, the most even distribution a rank-one
projector permits.

\begin{table}[h]
\caption{Ion/NMR qutrit, $T_1=1$~s and $T_2^*=5$~ms, record $2\ \mu$s. Ergotropy
error in units of the coherent contribution; readout fidelity $0.995$.}
\begin{ruledtabular}
\begin{tabular}{ccc}
shots & mean & s.d. \\
\hline
$1\times10^{4}$ & $+0.012$ & $0.126$ \\
$1\times10^{5}$ & $+0.004$ & $0.039$ \\
$1\times10^{6}$ & $+0.001$ & $0.013$ \\
\end{tabular}
\end{ruledtabular}
\end{table}

The noiseless bias is $[0,0,0]$ to machine precision with both $T_1$ and
$T_2$ channels active, and the eigen-operator residual vanishes exactly. The
ergotropy is $1.565$~MHz with a coherent share of $50.2\%$. Errors decrease as
$1/\sqrt N$ with no floor. Such a platform is the natural place to test
the immunity theorem at the $10^{-3}$ level; it is not the place to demonstrate
an advantage, since dephasing is negligible there anyway and the ruler's selling
point --- immunity to noise --- has nothing to bite on.

\subsection{S11.5 Individual (non-collective) relaxation}

The TC simulations above treat atomic decay through the collective jump operator
$\hat J_-=\sum_i\hat\sigma_i^-$, which maps the symmetric Dicke subspace to itself. A
real transmon relaxes through its own local channel
$\hat L^{(i)}=\sqrt{\gamma_1}\,\hat\sigma_i^-$, and a local jump does not preserve that
subspace. This subsection closes the gap by simulating the proposal on the full
excitation-number ladder instead of the symmetric chain.

\paragraph{Model.} With three transmons and one cavity mode at zero
temperature, let $\hat n=\hat a^\dagger\hat a$ be the photon-number operator,
$\hat K=\sum_i\hat\sigma_+^{(i)}\hat\sigma_-^{(i)}$ the number of atomic
excitations and $\hat N=\hat K+\hat n$ the total excitation number, so that $N$
denotes its eigenvalue. Then $\hat N$ is conserved by $\hat H_{\rm TC}$ and
lowered by every relaxation channel, so the ladder $N=2,1,0$ is closed. It contains
$|ggg,2\rangle$, the three states with one atomic excitation and one photon, and
the three with two excitations and no photon ($N=2$, seven states); the
corresponding $N=1$ sector (four states) and $N=0$ (one state), twelve in all.
Of these, the TC Hamiltonian drives only the three-dimensional symmetric chain
$|ggg,2\rangle\leftrightarrow|W,1\rangle\leftrightarrow|D_2,0\rangle$; the
remaining nine states are dark.

\paragraph{Two structural facts.} Both are verified numerically, not assumed.

(F1) The read-out operator $\hat O=\hat n$ has zero matrix elements between
the symmetric chain and any dark state of the same $N$ ($1.2\times10^{-16}$),
between the $N=1$ symmetric and $N=1$ dark states ($0$), and between different
$N$ ($0$). This is because $\hat O={\rm 1\!l}_{\rm at}\otimes\hat n_{\rm cav}$ and the
atomic factors are orthogonal. Inside a dark sector $\hat O$ has eigenvalues only
$0$ and $1$. Dark states therefore produce no spectral line at any
frequency; they can only shift the dc level of the record.

(F2) Relaxation populates the $N=1$ sector coherently, and that sector has its
own TC splitting $\sqrt{D^2+12g^2}=27.87$~MHz, which does appear in the record
as an extra line. It sits $1.52$~MHz from $f_{12}=26.36$~MHz. Its amplitude,
however, is more than an order of magnitude below the three chain lines, and
adding it to the fit basis changes the extracted coherences by less than $1\%$;
adding it in the presence of shot noise is mildly harmful, because two
basis functions $1.5$~MHz apart over a $2\ \mu$s record are strongly collinear
and inflate the variance. We therefore do not include it.

\paragraph{Criterion.} On the full ladder the eigen-operator residual is
computed exactly as in S11.2.

\begin{table}[h]
\caption{Eigen-operator residual on the full 12-state ladder, at
$T_1=20\ \mu$s, $T_2^*=30\ \mu$s and $\kappa^{-1}=3\ \mu$s. The ordering of the
channels matches Table~S11.2; absolute values differ from it because that table
used the three-state symmetric chain, whereas here the jump operators act on the
whole ladder.}
\begin{ruledtabular}
\begin{tabular}{lc}
Channel & Residual \\
\hline
Energy-basis dephasing & $9.8\times10^{-14}$ \\
Individual relaxation $\hat\sigma_i^-$ & $3.4\times10^{-4}$ \\
Collective relaxation $\hat J_-$ & $8.9\times10^{-4}$ \\
Individual dephasing $\hat\sigma_z^{(i)}$ & $3.7\times10^{-4}$ \\
Cavity decay $\sqrt\kappa\,\hat a$ & $2.3\times10^{-3}$ \\
\end{tabular}
\end{ruledtabular}
\end{table}

Individual relaxation violates the criterion less than cavity decay,
which was already shown in S11.3 to cost at most $\sim1\%$ of the coherent
ergotropy. Note also that individual relaxation violates it $2.7\times$ less
than the collective $\hat J_-$ used in the earlier simulations: the collective
approximation is conservative, because $\hat J_-$ has matrix elements $\sqrt3$
larger and hence drains the chain faster. No earlier conclusion needs revising.

\paragraph{Bias and its cure.} The record is fitted with the three true chain
frequencies; the second strategy adds a single linear term $\propto t$, which
requires no calibration and absorbs the population drift that (F1) says is the
only visible effect of the dark sectors.

\begin{table}[h]
\caption{Individual relaxation, noiseless, record $T=2\ \mu$s. Deviation is the
maximum relative error over the three coherences; the ergotropy error is in
units of the coherent contribution of the prepared state. Populations are
evaluated at $t=T$.}
\begin{ruledtabular}
\begin{tabular}{ccccccc}
 & \multicolumn{2}{c}{3 lines} & \multicolumn{2}{c}{3 lines $+t$}
 & \multicolumn{2}{c}{populations} \\
$T_1$ ($\mu$s) & dev & erg & dev & erg & $N{=}1$ dark & $N{<}2$ \\
\hline
10 & $4.3\times10^{-2}$ & $-0.003$ & $1.3\times10^{-2}$ & $-0.001$ & 0.038 & 0.182 \\
20 & $2.2\times10^{-2}$ & $-0.002$ & $6.6\times10^{-3}$ & $-0.000$ & 0.021 & 0.095 \\
50 & $8.6\times10^{-3}$ & $-0.001$ & $2.6\times10^{-3}$ & $-0.000$ & 0.009 & 0.039 \\
100 & $4.3\times10^{-3}$ & $-0.000$ & $1.3\times10^{-3}$ & $-0.000$ & 0.005 & 0.020 \\
200 & $2.1\times10^{-3}$ & $-0.000$ & $6.6\times10^{-4}$ & $-0.000$ & 0.002 & 0.010 \\
\end{tabular}
\end{ruledtabular}
\end{table}

The population in the $N=2$ dark sector is exactly zero at every $T_1$: a local
jump lowers the excitation number by one, so it transfers population from $N=2$
directly to $N=1$ and never into the dark states of the $N=2$ sector. The
$N=1$ dark states are populated (second-to-last column) but are invisible by
(F1). The bias scales as $\gamma_1T$, and the linear drift term removes about
two thirds of it; higher-order drift terms ($t^2,t^3,t^4$) give no further
improvement, the remainder being envelope mismatch rather than drift.

\paragraph{Comparison with the collective approximation.} At matched
single-atom rate, $\hat J_-$ gives a coherence deviation $4.0\times10^{-3}$ and an
ergotropy error $-0.000$ at $T_1=200\ \mu$s, versus $2.1\times10^{-3}$ and
$-0.000$ for the local channel; at $T_1=10\ \mu$s, $9.0\times10^{-2}$ versus
$4.3\times10^{-2}$. The collective model overestimates the error by a factor
$\approx2$ throughout, confirming that the numbers in S9 and S11.3 are
conservative.

\paragraph{With shot noise.} Repeating the Monte Carlo of Sec.~S9 with the full
realistic channel set --- individual $T_1=20\ \mu$s, local dephasing
$T_2^*=30\ \mu$s, $\kappa^{-1}=3\ \mu$s, readout fidelity $0.97$ --- gives an
ergotropy error of $0.216\pm0.358$ at $10^{5}$ shots, $0.051\pm0.160$ at
$3\times10^{5}$ and $0.000\pm0.071$ at $10^{6}$
(units: coherent contribution of the prepared state, whose absolute value is
$642$~MHz). The systematic error from individual relaxation, $0.002$, is
therefore two orders of magnitude below the statistical error at the shot budget
recommended in Sec.~S9. Individual relaxation is not the bottleneck, and no
change of platform is required to deal with it.

\begin{figure}[h]
\includegraphics[width=\textwidth]{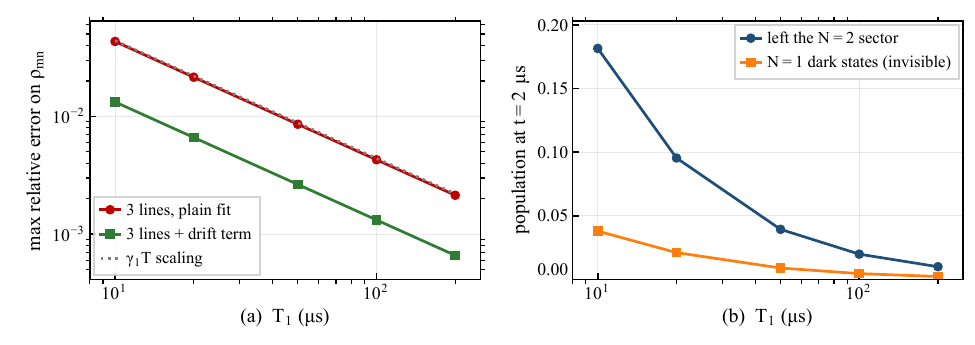}
\caption{(a) Maximum relative error on the extracted coherences versus $T_1$,
with and without the linear drift term; the dotted line is the $\gamma_1T$
scaling. (b) Population leaving the $N=2$ sector during the $2\ \mu$s record,
split into the visible cascade and the invisible $N=1$ dark states.}
\end{figure}

\subsection{S11.6 What remains open}

The individual-relaxation gap is closed above; one gap is not. The criterion
Eq.~\eqref{eq:jumpterm} is stated per jump operator and is both necessary
and sufficient for a single $\hat L$ (with non-degenerate $\hat H$), so it is not
merely a convenient sufficient condition. What remains open is the multi-channel
case: the sum $\sum_\alpha \hat L_\alpha^\dagger \hat A_{mn}\hat L_\alpha$ could in principle
have its off-diagonal parts cancel between different $\alpha$, in which case a
model excluded by the per-operator criterion would still be immune. We have not
found such a cancellation, but we do not exclude it. Neither is the effect of
a thermal (rather than zero-temperature) bath, which adds excitation channels
$\hat L_i=\sqrt{\gamma_\uparrow}\,\hat\sigma_i^+$ and would populate the ladder from
below; at $20$~mK and $6$~GHz the thermal occupation is $\lesssim10^{-5}$ and we
have neglected it.

\bibliography{qdrs}